\documentclass[preprint,authoryear,12pt]{elsarticle}

\usepackage{amssymb}
\usepackage[normal]{subfigure}

\usepackage{lineno}

\journal{Icarus}

\begin{document}

\begin{frontmatter}



\title{Bistability of the climate around the habitable zone: a thermodynamic investigation}


\author[label1]{Robert Boschi}
\author[label1,label2]{Valerio Lucarini}
\author[label1]{Salvatore Pascale}

\address[label1]{Klimacampus, Meteorologisches Institut, Universit\"at  Hamburg, Hamburg, Germany}
\address[label2]{Department of Mathematics and Statistics, University of Reading, Reading, UK}

\begin{abstract}
The goal of this paper is to explore the potential multistability  of the climate of a planet around the habitable zone. We apply our methodology to the Earth system, but our investigation has more general relevance. A thorough investigation of the thermodynamics of the climate system is performed  for very  diverse conditions of energy input and infrared atmosphere opacity.
 Using PlaSim, an Earth-like general circulation model, the solar constant $S^*$ is modulated between $1160$ and $1510$ Wm$^{-2}$ and   the CO$_2$ concentration, [CO$_2$],  from $90$  to $2880$ ppm. It is observed that in such a parameter range   the climate is bistable, i.e. there are two coexisting attractors, one characterised by warm, moist climates (W) and one by completely frozen sea surface (Snowball Earth, SB). Linear relationships  are  found for the two transition lines (W$\rightarrow$SB and SB $\rightarrow$W) in ($S^*$,[CO$_2$]) between $S^*$ and the logarithm of [CO$_2$]. The dynamical and thermodynamical properties -- energy fluxes, Lorenz energy cycle, Carnot efficiency, material entropy production -- of the  W and SB states are very different: W states are dominated by the hydrological cycle and latent heat is prominent in the material entropy production; the SB states are eminently dry climates where heat transport is realized  through sensible heat fluxes and entropy mostly generated by dissipation of kinetic energy. We also show that the Carnot efficiency regularly increases  towards each transition between W and SB, with a large decrease   in  each transition. Finally, we propose well-defined empirical functions allowing for expressing the global non-equilibrium thermodynamical properties of the system in terms of either the mean surface temperature or the mean planetary emission temperature. This paves the way for the possibility of  proposing efficient parametrisations of complex non-equilibrium properties and of practically deducing fundamental properties of a planetary system from a relatively simple observable.

\end{abstract}


\end{frontmatter}

 \linenumbers


\section{Introduction}

\subsection{Planetary atmospheres and extrasolar planets} 
\label{intro1}

A very active recent  field of research in astrophysical sciences is the observational, model-assisted, and theoretical investigation of extra-solar planetary objects. In  less than two decades, the development of new instruments have lead to the detection of the first extra-solar planet in the mid 90Õs to more and more refined observations of the characteristics of several hundreds of such bodies. The properties of planets vary greatly, in terms of their compositions  - gaseous or rocky -, nature of their atmosphere, and of their size - ranging in several orders of magnitude. Astronomical and astrophysical factors of great relevance include the temperature and the intensity of the radiation emitted by the parent star, the orbital parameters, and whether or not the planet is tidally locked. A great deal of effort has been directed at constraining the combinations of physical configurations potentially compatible with life (habitable zone) , that is to say  with the possibility of observing water prominently or at least partially in its liquid form at surface. Obviously, the so-called habitable zone is the setting where we can hope to find forms of extraterrestrial life, at least assuming that life necessarily expresses itself in forms analogous to those we experience on our planet.  Recently, ESOÕs HARPS planet finder estimated that just in the Milky Way billions of habitable Earth-like rocky planets could orbit around faint red dwarfs,with in the order of one hundred in the immediate vicinity of our Solar system \citep{Bonfils}.  We refer the reader to the website (http://exoplanet.eu/index.php) --  which is dedicated to collecting information on all newly discovered planets and on the related bibliography -- and  recently published  books \citep{Dvorak, Saeger, Kasting, Perryman}. 

A great deal of interest is now being paid to inferring, studying, and the modeling of planetary atmospheres, i.e., in the case of rocky planets, the fluid envelope that surrounds the planet and the response to the differential heating due to the absorption of the incoming radiation. Planetary atmospheres may feature complex chemical compositions and be characterised by the relevant presence of phase transitions of some of the components, which may have large effects on the radiative properties. The geometry describing how the planet and its parent  star face each other determines to a first approximation the horizontal inhomogeneities, whereas the atmospheric optical properties, and in particular its opacity to the incoming radiation, has an effect on the vertical structure (e.g.  Earth's the optically thin atmosphere is   typically heated from the  planets surface). The general circulation of the planetary atmosphere is powered, just like a Carnot engine by such inhomogeneities with, on average, heating occurring at higher temperatures being then removed in cooler regions. The circulation acts to reduce the temperature gradients through  transport  processes, and on average the generation of kinetic energy is compensated by its irreversible dissipation through viscous processes. Irreversible heat transport and dissipation, and  processes of radiation absorption and emission contribute to the generation of entropy. The steady state is realised by  balancing energy and entropy fluxes between the planetary atmosphere and the surrounding space.  Various aspects of this closed chain of interdependent processes have been described for terrestrial conditions  \citep{Lor67,Peix2,HeldSoden,Lucarini,Luc10}, but these arguments could be applied to planetary atmospheres more generally.

The recent  discoveries in planetary astronomy and astrophysics are progressively affecting classical geophysical sciences (by definition Earth-centered). The study of, e.g. JupiterÕs or VenusÕ general circulation has a rather old history, but with the recent and foreseeable future discoveries of exoplanets we have the opportunity of facing a vast variety of planetary configurations, so that we are going from considering special cases of climates (Earth, Jupiter, Saturn, etc), to being able to study a quasi-continuous distribution of climates in some parametric space. Following \cite{Read}, one should note that it is possible to reduce the variety of climate settings by adopting the fluid-dynamical classical method of similarity, i.e. by defining a set of dimensionless  numbers that fully characterise the climate states. When two climate states share the same set of dimensionless numbers, they are dynamically equivalent, so that the statistical properties of one can be mapped into those of the other one with simple algebraic operations.
At this stage, we are left with two additional elements to cope with:
\begin{itemize}
\item 	how to characterise succinctly a climatic state, conveying minimal but comprehensive physical information: the planetary atmospheres are, in general, turbulent fluids with variability on a very vast of range of spatial and temporal scales?
\item          how to verify, at least approximately, the validity of our simulations and of our theoretical understanding of the planetary circulations we are able to model?
\end{itemize}

The two items are closely related to the definition of robust observables. Energy conservation imposes that the incoming radiation onto a planet is instantaneously equal to the sum of the radiation that is absorbed and scattered by the planet. Assuming steady state, when averaging over  a sufficiently long time  interval, the radiation absorbed by the planet is equal to the radiation the planet emits off to space. These quantities allow defining the average albedo of the planet and its effective thermodynamic temperature, which constitute the most fundamental description of the properties of the planet in terms of the first law of thermodynamics. The observation of a planet as represented by a single pixel recorded by a telescope permits to gather information at this level of detail in terms of its overall macroscopic properties. Obviously, the observations provide much more than this, such as the spectrum emitted by the planet (and by the star). Nonetheless, in terms of the macroscopic thermodynamic properties of the planet, only the spectrally integrated quantities are mostly  relevant.
Under conditions of steady state, given the inhomogeneity of the incoming radiation and of the boundary conditions, the vanishing net energy budget at the top of the atmosphere (TOA) results from the cancellation between regions where the budget is positive and regions where the budget is negative (low and mid-high latitudes regions, in the case of the Earth). Steady state argument imply that the absolute value of the imbalance in either region is equal to the energy transmitted from the regions where the energy balance at the top of the atmosphere is positive from those where such balance is negative, and the transport is performed across the atmospheric fluid envelope through material transport and horizontal deflection of the incoming radiation (in the case of very thick atmospheres). Note that the correct representation of these large scale properties in state-of-the-art models   of the Earth's climate  is far from being a trivial task \citep{LucariniRagone}.

Moreover, the second law of thermodynamics imposes that the regions featuring positive TOA energy budget are warmer (i.e. their total emitted radiation is larger) than those where the balance is negative. One can show  that the entropy production due to such irreversible transport of energy from hot to cold regions gives a lower bound to the total material entropy production of the planet. Moreover, it is possible to provide a lower bound to the total dissipation of kinetic energy  or, equivalently, to the intensity of the Lorenz energy cycle \citep{Luc10}. This discussion implies that if a telescope allows for an observational resolution involving more than one pixel, able to distinguish between warmer and colder regions, it is possible to derive fundamental  information on the macroscopic thermodynamic properties of the planet in terms of the 2nd law of thermodynamics.  This brief discussion suggests that non-equilibrium thermodynamics provides rather powerful methods and concepts for analysing the fundamental properties of planetary atmospheres, for testing modelsÕ performances, and for deriving bounds on the physical properties of the system even when low-resolution data are available. Following previous work in the field \citep{Puj03, Frae, Lucarini, LucFr, Luc10, Pascale, Pascale2, Jiangnan}, we propose to use this framework in order to study planetary atmospheres in a rather general setting of forcings and boundary conditions. 

\subsection{Habitability conditions and climatic bistability}
\label{intro2}

As mentioned above, a great deal of interest in the investigation of planetary atmospheres is directed at studying conditions within or about the Habitable zone. Nonetheless, being in the the Habitable zone is a \emph{necessary but  not a sufficient condition} for a planet to have liquid water at the surface, even if the chemical composition of the atmosphere would in principle allow for it. In fact, the paleohistory of our planet provides strong evidence of the fact that the astronomical and astrophysical parameters of the Sun-Earth system support two distinct steady states, the warm state (W)  characterised by widespread liquid water, and one, instead, characterised by the global glaciation of water and an extremely dry atmosphere, i.e. the so-called Snowball state (SB). What we propose here is a thorough investigation of these two states in a bidimensional parameteric space, with the goal of detailing the region of bistability of the Earth system and analysing the bifurcation points and the mechanisms of transitions between the two states.  Our investigation deals with Earth conditions because this is the only system where an extensive body of literature Ð observational, theoretical, amd model-assisted --  on the Snowball state and the Snowball-Snowfree transitions is available, but our scope is much wider.

Probably the most notable examples of climate change events occurred during the Neoproterozoic (period spanning from 1000 million to 540 million years ago), when the Earth is believed to have suffered two of its most severe periods of glaciation \citep{Hoffman} and entered into  a SB climate state.  This period coincided with large carbon dioxide fluctuations, while the solar constant (about 1365 Wm-2 in present conditions) is believed to have been 94 $\%$ of current levels, rising to 95$\%$  by the end of the Neoproterozoic  \citep{Gough,Pierre11}. The two main factors effecting concentration of atmospheric CO$_2$ are  biotic activity and volcanism. Volcanic eruptions bring about very sudden and dramatic increases in CO$_2$ concentration. This process  provides a potential mechanism through which      the climate state can exit the SB condition,  by increasing the opacity of the atmosphere and enhancing the greenhouse effect. The biospheric effect tends to occur more gradually as the biotic activity and atmospheric composition are coupled so that large fluctuations of the carbon pools take place over large time scales. Note that the effect of SB events on the biosphere is believed to have been disastrous. Carbon-isotope ratios characteristic of EarthÕs mantle \citep{Hoffman, Kennedy}  rather than of life processes  recorded immediately below and above the glacial deposits  imply that oceanic photosynthesis was effectively non-existent during SB events. The result of this and anoxic conditions beneath the ice should have lead to the disappearance of most kinds of forms of life except bacteria. The final disappearance of SB conditions since the Neoproterozoic may have been the main contributing factor in the development of complex multi-cellular life that began around 565 million years ago.

Based on the evidence supported by \citep{Hoffman, HoffmanSchrag}, it is therefore expected that the Earth is potentially capable of supporting multiple steady states for the same values of some parameters such as  the solar constant and the concentration of carbon dioxide, which directly affect the radiative forcing. It is important therefore to explore this hypothesis, due to the relevance for the history of our planet but also to help understand other planets capability for supporting life.
Initial research using simple $0$-D models \citep{Budyko,Sellers}, $1$-D models \citep{Ghil} as well as more recent analyses performed using complex 3-D general circulation models \citep{Maro,Voigt,Pierre11},  provide support for the existence of such bistability. The SB$\rightarrow$W and W$\rightarrow$SB transitions tend to occur in an abrupt rather than a smooth transition. The main mechanism triggering such abrupt transitions is the positive ice-albedo feedback \citep{Budyko,Sellers}. Such a feedback is associated with the fact that as temperatures increase, the extent of snow and ice cover decreases thus reducing the albedo and therefore increasing the amount of sunlight absorbed by the Earth system. Conversely, a negative fluctuation in the temperature  leads to an increase in the albedo therefore reinforcing the cooling.
  The presence of such catastrophic climate shifts \citep{Arnold} suggest the existence of a global bifurcation in the climate system for certain combinations of its descriptive parameters \citep{Frae79}.  The loss of stability realized in the W$\rightarrow$SB and SB$\rightarrow$W transitions is related to the catastrophic disappearance of one of the two attractors describing the two possible climatic states, as a result of a set of complicated bifurcations.

          Starting from present conditions, the most obvious physical parameters to modulate in order to bring about the transition to the SB state is  the solar constant.  Even if other model experiments \citep{Voigt} show that the decrease in CO$_2$ alone can bring about transition to the SB state, this requires approximately an 80$\%$  decrease in CO$_2$ concentration, compared to a decrease of less than 10$\%$ for the solar luminosity. The Neoproterozoic however highlights the importance of considering the changes in the CO$_2$ levels as a mechanism for the transitions to and from the SB climate state, and therefore the dramatic impact it can have on the overall state of the climate system. It is therefore interesting to alter both the solar luminosity and the atmospheric opacity as these are two important parameters affecting the overall properties of the system. If one wants to explore extensively the parametric space of climate steady states, it is therefore necessary to consider a wide range of values for both of these parameters.

 Using PlaSim, a general circulation model of intermediate complexity \citep{Frae2}, we study the climate states realised when the solar constant is modulated between $1160$ Wm$^{-2}$  and $1510$ Wm$^{-2}$ and the  values of [CO$_2$] are varied beetwen  90 to 2880 ppm.  Our aim here is an empirical (i.e. based on model simulations) reconstruction of the global structural properties of the climatic attractors. For both W and SB states we compute surface temperature, material entropy production, meridional energy transport, Carnot efficiency \citep{Johnson} and dissipation of kinetic energy and propose empirical relationships in the parametric plane ($S^*$,[CO$_2$]). We will look for an empirical relation for the two transition lines (W$\rightarrow$SB and SB$\rightarrow$W) in the parametric plane between $S^*$ and the natural logarithm of [CO$_2$] which marks the boundaries of the hysteresis in the climate system. The aforementioned quantities are  used to explain changes in large-scale climate behaviour and the effect of climate change on features such as stratification and baroclinicity in order to understand changes in the meridional heat transport across the parameter range. It will also be shown that the Carnot efficiency has a key role in defining the stability of the system which is related to abrupt climatic shifts.
 
The paper is structured in the following way: in Section~\ref{Noneq} we describe  details of the non-equilibrium thermodynamics of the climate and the diagnostic tools used. Section~\ref{Exp} is dedicated to the description of the PlaSim climate model and of the experimental setup.  In Section~\ref{Results}  we discuss the results of the simulations, in Section~\ref{param} we propose parametrisations of the main non-equilibrium properties of the system as a function of the mean surface temperature or emission temperature,    and in Section~\ref{concl} we present our conclusions.

\section{Non-equilibrium Thermodynamics of the climate}
\label{Noneq}

In this section we recapitulate some thermodynamic properties of the climate system and introduce the notation used throughout this chapter. We follow what presented in \cite{Lucarini}.  If the climate system is encompassed by a domain $\Omega$, the total energy budget is given by $E(\Omega) = P(\Omega) + K(\Omega)$, where $K$ represents the total kinetic energy and $P$ is the moist static potential energy, which includes contributions from the thermal (including latent heat) and potential energy. The time derivative of $K$ and $P$ can be found to be $\dot{K}=-D+W$ and $\dot{P}=\dot{\Psi}+D-W$,  where $ D$ is the dissipation and therefore always positive, $W$ is the instantaneous work done by the system and $\dot{\Psi}$ which is the heating due to convergence of turbulent heat fluxes and radiative heat, such that $\dot{E}=\dot{\Psi}$. The dependence on $\Omega$ has been dropped for convenience.   As soon as the climate is considered as a non-equilibrium steady state system (NESS, see \cite{Gallavotti}), we have that over long time scales $\overline{\dot{E}}=\overline{\dot{P}}=\overline{\dot{K}}=0$  (the bar indicates averaging over long time periods).   Let us define $\dot{Q}$   as the local heating rate so that $\dot{Q}=\rho(\epsilon^2-\nabla\cdot H$) \citep{Lucarini} where $\epsilon^2 >0$ is the local rate of heating due to viscous dissipation of kinetic energy and  $H$   is given by the sum of turbulent heat fluxes plus radiative energy fluxes. We split the domain $\Omega$ into the domain $\Omega^+$ , where $ \dot{Q}=\dot{Q}^+ >0$,   and $\Omega^-$   in which $\dot{Q}=\dot{Q}^+ <0$.  Therefore, we find that  $\dot{Q}^+$  and $\dot{Q}^-$ integrated over $\Omega$ equal the derivative of the total heating due to dissipation, $D$  and the convergence of heat fluxes $\dot{\Psi}$:

\begin{equation}
\dot{\Psi}+D=\dot{P}+W=\int_{\Omega^+} dV \rho Q^+  + \int_{\Omega^-} dV \rho  Q^- = \dot{\Phi}^+ + \dot{\Phi}^- = \dot{\Phi}
\end{equation}

Where the quantities $\dot{\Phi}^+$ and $\dot{\Phi}^-$  are positive and negative at all times, respectively. Since dissipation is positive definite, $-\overline{\dot{K}}+\overline{W}=\overline{D}=\overline{\dot{P}}+\overline{W}=\overline{W}=\overline{\dot{\Phi}^+} + \overline{\dot{\Phi}^- }> 0$.
	On spatial scales far smaller than $\Omega$ itself, it is practical to assume local equilibrium (local thermodynamic equilibrium hypothesis, \cite{Maz}) so that locally $ \dot{Q}=\dot{s} T $ with $\dot{s}$  the time derivative of the entropy density. The total rate of change of the entropy of the system is:

\begin{equation}
\dot{S}=\int_{\Omega^+} dV \rho  \dot{Q}^+ /T + \int_{\Omega^-} dV \rho \dot{Q}^- /T= \int_{\Omega^+}    dV \rho |s^+ | + \int_{\Omega^-} dV \rho |\dot{s}^- | = \dot{\Sigma}^+ + \dot{\Sigma}^-  
\label{Srate}
\end{equation}

where $\dot{\Sigma}^+  >0$ and $\dot{\Sigma}^- <0$. Using equation (2) and assuming that the Earth system is in a steady state, over a long time average, $\overline{\dot{\Sigma}^+ }=-\overline{\dot{\Sigma}^- }$ as $\overline{\dot{S}}=0$. Therefore, $2 \overline{\dot{\Sigma}^+} =\overline{\int_{\Omega} dV \rho |\dot{s}|} $, so that  $\overline{\dot{\Sigma}^+ }$ measures the absolute value of the entropy fluctuations throughout the domain.
When integrating over the whole domain and considering long time average, we have the following equivalent expressions for the thermodynamic quantites:  $\overline{\dot{\Phi}^+ }=\overline{\dot{\Sigma}^+}   \Theta^+$ and  $\overline{\dot{\Phi}^-} = \overline{\dot{\Sigma}^-}   \Theta^-$ , where $\Theta^+$  and $\Theta^-$  are the time and space averaged temperatures of the $\Omega^+$ and $\Omega^-$ domains respectively. Since $|\dot{\Sigma}^+  |=|\dot{\Sigma}^- |$  and $|\Phi^+|>|\Phi^- |$, it can be shown that $\Theta^+ > \Theta^-$, i.e absorption typically occurs at higher temperature than release of heat \citep{Peix, Johnson}. The work done by the Carnot engine of the climate system is found to be, $\overline{W}=\eta \overline{\dot{\Phi}^+} $, where 
\begin{equation}
\eta=\frac{\overline{\dot{\Phi^+ }}  + \overline{\dot{\Phi^-}  }}{\overline{\dot{\Phi^+}}}    
\end{equation}
can be defined as the Carnot efficiency of the system.
   As shown in Lorenz \cite{Lor67} Ð and clarified in \cite{Johnson} (2000) Ð the long term average of the work performed by the system is equal to the long-term average of the generation of available potential energy, as typical of forced-dissipative steady states. The Earth exists in a steady state maintained far from equilibrium by net radiative heating in the warm region (low latitude in our planet) and net cooling at the cold regions (high latitude in our planet),  which are compensated by large scale transports performed by the planetary atmpsphere. This gives rise to ongoing irreversible processes, including phase transitions  (H$_2$O  in the case of our planet) and frictional dissipation, which are characterized by a positive entropy production. The entropy production due to the irreversibility of the processes occurring within the climatic fluid is called the material entropy production, $\dot{S}_{mat}$ and  can be written in general terms as:
\begin{equation}
\overline{\dot{S}_{mat} }= \overline{\int_{\Omega} \frac{\epsilon^2}{T} dV} +   \overline{\int_{\Omega}\vec{F}_{SH} \cdot\nabla\frac{ 1}{T} dV}+\overline{\int_{\Omega}  \vec{F}_{LH} \cdot\nabla\frac{ 1}{T} dV}    
\label{smater}
\end{equation}
where the first, second, and third terms on the RHS are related to  the dissipation of kinetic energy, and to the transport of sensible and latent heat  respectively. We now wish to link the terms of the entropy budget in eq. (\ref{Srate}) with those of the entropy production in eq. (\ref{smater}).
The second law of thermodynamics states that the entropy variation of a system at temperature $T $ receiving an amount of heat $\delta Q$ is larger than or of equal to $\delta Q/T$ \citet{Landau}. In this case:
\[
\overline{\dot{S}_{mat} (\Omega)} \ge \overline{\dot{S}_{min} (\Omega)}=  \overline{ \left( \frac{\int_{\Omega}dV \rho \dot{Q})}{\int_{\Omega}dV \rho T} \right) }  = \overline{ \left( \frac{\Phi^+ + \Phi^-}{\langle\Theta\rangle } \right)}  \approx 
\]
\begin{equation}
\approx    \frac{\overline{\Phi^+} +\overline{\Phi^- }}{\overline{\langle \Theta \rangle }}   \approx  \frac{\overline{\Phi^+} + \overline{\Phi^- }}{\frac{(\Theta^+   +\Theta^- )}{2}}= \frac{\overline{W}}{\frac{(\Theta^+ +\Theta^- )}{2}} 
\end{equation}
where $\overline{\dot{S}_{mat} (\Omega)}$ is the long-term average of the material entropy production, $\overline{\dot{S}_{min} (\Omega)}$ is its lower bound, i.e. the minimal value of the entropy production compatible with the presence of a Lorenz energy cycle with average intensity $W?  ?$ and $\langle \Theta \rangle$ is the density  averaged temperature of the system. The approximation holds as long as we can neglect the impact of the cross-correlation between the total net heat balance and the average temperature and we can assume that  $\langle  \Theta \rangle $  can be approximated by the mean of the two Carnot temperatures $\Theta^+$ and $\Theta^-$. We can explicitly write $\dot{S}_{min} (\Omega)$ as:
\[
\overline{\dot{S}_{min} (\Omega)} \approx  \frac{\overline{W}}{(\Theta^+ +\Theta^- )/2}=  \frac{\eta \overline{\Phi^+ }}{(\Theta^+  +\Theta^- )/2} =
\]
\begin{equation}
=\eta \frac{\Theta^+ }{(\Theta^+ + \Theta^- )/2} \overline{\dot{\Sigma}^+ }=\frac{\eta}{1-\eta/2} \overline{\dot{\Sigma}^+ }\approx \eta\overline{\dot{\Sigma}^+}    
\end{equation}
where the last approximation holds as long as $\eta \ll 1$, which is typically the case.  Therefore, $\eta$ sets also the proportionality factor relating the lower bound to the entropy production of the system $\overline{\dot{S}_{min} (\Omega)}$ Ð due to macroscopically irreversible processes Ð to the absolute value of the entropy fluctuations inside the system due to macroscopically reversible heating or cooling processes. Note that if the system is isothermal and at equilibrium the internal entropy production is zero, since $\eta\rightarrow0$. The lower bound to the material entropy production corresponds to the contribution coming from the dissipation of kinetic energy through viscous processes. Therefore, the average material entropy production can be expressed as $\overline{\dot{S}_{mat}} =\overline{\dot{S}_{min}} +\overline{\dot{S}_{exc}} $, where $\dot{S}_{exc} $ is the excess of entropy production with respect to the minimum, which results from the heat transport down the temperature gradient (\citep{Lucarini}). We can define:
\begin{equation}
 \alpha \approx \frac{\overline{S_{exc}}}{\overline{S_{min}}}  \approx \frac{\int_{\Omega} dV \overline{(\vec{F}_{SH}+\vec{F}_{LH})\cdot\nabla(\frac{1}{T}) } }{(\overline{W}/\langle \Theta \rangle) } \ge 0           
 \end{equation}
as a parameter of the irreversibility of the system, which is zero if all the production of entropy is due to the Ð unavoidable Ð viscous dissipation of the mechanical energy. As $\overline{ \dot{S}_{mat}} \approx \eta \overline{\dot{\Sigma}^+} (1+\alpha)$, we have that the entropy production is maximized if we have a joint optimization of heat transport and the production of mechanical work. Note that, if heat transport down the temperature gradient is very strong, the efficiency $\eta$ is small because the difference between the temperatures of the warm and cold reservoirs is greatly reduced (the system is almost isothermal), whereas, if the transport is very weak, the factor $\alpha$ is small.  The parameter $\alpha$ introduced above is related to the Bejan number $\mathcal{B}e$ as $\mathcal{B}e=\alpha+1$ \citep{Paoletti}.

\section{Experimental setup}
\label{Exp}

We study dynamical and thermodynamical properties  of the EarthÕs  climate  using the PlaSim climate model.
This is motivated by the fact that our study requires altering very extensively some parameters of the climate system and producing many simulations. Therefore we need a model which is flexible and fast to run rather then a state-of-the-art model encompassing as many processes occurring in the Earth as possible.
 PlaSim \citep{Frae2} is a climate model of intermediate complexity, freely available at htpp://www.mi.uni-hamburg.de/plasim.  Its dynamical core is formulated using the primitive equations for vorticity, divergence, temperature and the logarithm of surface pressure, solved using the spectral transform method \citep{Eliasen,Orszag}. Unresolved processes for long \citep{Sasamori} and short \citep{Lacis} wave radiation, shallow, moist  \citep{Kuo, Kuo2} and dry convection, cloud formation \citep{Stephens,Stephens2,SlingoSlingo} and large scale precipitation, latent and sensible heat boundary layer fluxes, horizontal and vertical diffusion  \citep{Louis,Louis2,Laursen}  are parameterized. The model is coupled to a 50-$m$ deep mixed layer ocean which contains a thermodynamic sea-ice model. The advantage of using a slab ocean as opposed to a full ocean is that it allows for the climate system to reach a stead state in less than 35 years after a change in e.g. the solar constant. With full ocean coupling, the integration time of the model and the time needed to reach a steady state would be an order of magnitude larger \citep{Voigt}.  We wish to emphasize that whereas most state-of-the-art general circulation models feature considerable energy imbalances, as highlighted by \cite{Luc10}, the energy bias is of the order of  $0.5$ Wm$^{-2}$ in very extreme climate conductions and less than that for more usual choices of climatic parameters. Moreover a tested  entropy diagnostic is available \citep{Frae}, thus making it well suited for this work.
The model is run at T21 resolution (approximately $5.6^{\circ}\times 5.6^{\circ}$) with 10 vertical levels. Changing $S^*$ over a wide range of [CO$_2$] of 90, 180, 270, 360, 540, 720, 1080, 1440, 2160 and 2880 
ppm, we are able to reconstruct the SB and W climate states. The procedure occurs as follows for each of the considered values of [CO$_2$]:
\begin{itemize}
\item[1.]	the model is run to a W steady state for 100 years with $S^*$ equal to 1510 Wm$^{-2}$;
\item[2.]	$S^*$ is decreased by a small amount for each value of CO$_2$ and the model run is continued until a steady state is reached;
\item[3.]	step 2 is repeated until $S^*$ is reduced to 1165 Wm$^{-2}$; the point of   W$\rightarrow$SB transition  is noted down;
\item[4.]	the reverse operation is then performed with $S^*$ increased step by step,  up to the value of   1510 Wm$^{-2}$, each time allowing the system to reach a steady state; the point of SB$\rightarrow$W  transition is noted down.
\end{itemize}
Further to this, we identify the position of the transition to a higher resolution than the rest of the parameter range in the direction of $S^*$ . For values of $S^*$ within $10$ Wm$^{-2}$ before the transition, $S^*$ is decreased in intervals of 1 Wm$^{-2}$, each time permitting 50 years for the system to reach a steady state, until after the transition is observed.

\section{Hysteresis, bistability and regime boundaries in a parametric space}
\label{Results}

\subsection{Temperature and entropy production}

Initially, the focus is put on analysing the parametric plane ([CO$_2$], $S^*$), which in the following shall be referred to as the CS space, as a function of global mean surface temperature, $T_s$. The transition zones between the main climate states are clearly  defined from the temperature profile. Note, the qualitative properties of the climate system in the CS space, namely the presence of bistability or of just one of the SB/W states, can be reconstructed from any observable of the climate state, but it is most instructive to select first the surface temperature because it is also practically the  most relevant. The temperature profile through the CS space is illustrated in Figures \ref{secondo_a} and \ref{secondo_b}.
We can identify two main climatic  regimes, observed as two distinct manifolds  ([CO$_2$], $S^*$, $T_s$) and characterized by a sharp change in the profile of $T_s$ when jumping from one manifold to another. We refer to these as the upper and lower manifolds, representative of the W and SB regimes respectively. As would be expected, there is a monotonic increase of temperature with increasing CO$_2$ or $S^*$ on both manifolds \citet{Voigt,Pierre05}. The temperature range on the SB and W manifolds are $212$ K-$242$ K, and $254$ K-$326$ K respectively, over the parametric space. We see that the temperature range $242$ K-$254$ K is not permitted by the climate system.  Note, due to the different temperature ranges the colour scaling of figures \ref{primo_a} and \ref{primo_b} is a factor of 4 different, with both scales starting from the same lowest value.

  The temperature range of the bistable region in the SB and W regimes are $218$ K-$242$ K and $254$ K-$300$ K respectively, meaning that the rate of change of surface temperature over the same range of $S^*$ and [CO$_2$] is approximately twice as large in the W regime  with respect to the SB regime, and that the surface temperature difference between the two manifolds ranges between $40$ K and $ 60$  K for identical values of $S^*$ and [CO$_2$]. The W states (upper manifold) exists only in the region of the CS space above the W$\rightarrow$SB transition line (the position of this line is expressed as $S^*=S_{sbw}^*$) whereas the SB states (lower manifold) only in the CS region below the SB$\rightarrow$W transition line (the position of this line is expresses as $S=S_{wsb}^*$). Such lines, which are well separated and approximately parallel, are illustrated as solid and dashed purple dashed lines on figures \ref{primo_a}-\ref{ottavo_b} and have been found within an accuracy of 2 Wm$^{-2}$ of the solar constant.  The solid purple lines indicate the ÔactiveÕ transition, dependant on which manifold the climate system lies in at that moment i.e the active transition for the SB and W states are SB$\rightarrow$W and W$\rightarrow$SB respectively. The dashed lines illustrate the location of the ÔinactiveÕ transition, when the climate system exists in the alternative state. The bistable region is therefore located between the dashed and solid purple lines.  As a result, a property of the system is that regardless of which combination of [CO$_2$ ] and $S^*$ is used, the transition from one state to another always occurs at almost exactly the same temperature. This indicates that the climate system is ``blind"  to the mechanism of forcing. 
The position of the two boundaries can be parameterised in terms of $S^*$ and CO$_2$ concentration as: 
\begin{equation}
S_{sbw}^* = a_{sbw}\log_{10}[CO_2] + C_{sbw} , \quad     S_{wsb}^* = a_{wsb}\log_{10} [CO_2] + C_{wsb}     
\end{equation}
where $a_{sbw}  \approx  a_{wsb} \approx  -70$ Wm$^{-2}$, $C_{sbw} \approx  1630$ Wm$^{-2}$ and $C_{wsb} \approx 1440$ Wm$^{-2}$ for the transition SB$\rightarrow$W and W$\rightarrow$SB respectively and  [CO$_2$] is expressed in ppm. The size of the bistable region, which we define as $B$,  along $S^*$ can therefore be defined by the difference between $C_{sbw}$ and $C_{wsb}$ :
\begin{equation}
B = C_{sbw}-C_{wsb}.      
\end{equation}
It is found that $B$ is  approximately $200$ Wm$^{-2}$.  The displacement between the position of the boundaries gives a precise measure of the hysteretic properties of the climate \citep{Budyko,Sellers,Voigt, LucFr}  since it indicates the size of the overlap between the two manifolds in the CS plane. 
The presence of a bistable region implies that when we change the values of $S^*$  and [CO$_2$] from an initial to a final value, the final steady state depends on the initial steady state and on the path of change of $S^*$ and [CO$_2$]. 
Let  us assume that we start from an initial  point $(S_0^*, [ CO_2 ]_0)$ in the bistable region and in the W state. Let us also assume we perform a close path of variation of $S^*$ and $[CO_2]$ so that $S_0^* = S_f^*$  and $[CO_2]_0  = [CO_2 ]_f$ . If the path does not cross  $S^*_{wsb}$ the final state will be identical to the initial one, that is, in a time-averaged sense:
\begin{equation}
T_s (S_0^*,[CO_2 ]_0)=T_s (S_f^*,[CO_2 ]_f).      
\end{equation}
On the other hand, if the closed path crosses the transition line to the second manifold, the final state will be different from the initial:
\begin{equation}
T_s (S_0^*,[CO_2]_0) \ne T_s (S_f^*,[CO_2 ]_f) .         
\end{equation}
If, furthermore, the closed path crosses first $S^*_{wsb}$  and then $S^*_{sbw}$, then again  $T_S(S_0^*,[CO_2]_0)=T_S (S_f^*,[CO_2]_f)$, since the system has performed first a W$\rightarrow$SB, and then a SB$\rightarrow$W transition. The same applies starting from a SB state and exchanging SB with W in the previous discussion. This is true for any climate diagnostic. More specifically, in the case of $T_s$,  for W$\rightarrow$SB and SB$\rightarrow$W transitions,
\begin{equation}
T_s (S_0^*,[CO_2 ]_0)>T_s (S_f^*,[CO_2 ]_f)  \quad \quad \textrm{and}  \quad \quad  T_s (S_0^*,[CO_2]_0)<T_s (S_f^*,[CO_2 ]_f)   
\end{equation}
respectively. Note that for $S^* > 1440$ Wm$^{-2}$, even if CO$_2$ is $ 0$ ppm, no transition to SB state can occur.  
        
          Figures \ref{terzo_a} and \ref{terzo_b} show the reconstruction for the material entropy production, $\dot{S}_{mat}$  in CS space, computed directly as described in \cite{Frae}. As with temperature, $\dot{S}_{mat}$ increases monotonically with increasing $S^*$ and $[CO_2]$  on both manifolds. In the SB state, the entropy is mostly generated by dissipation of kinetic energy and by irreversible 
sensible heat transport, because the planet is almost entirely dry. For the W manifold the main contribution to entropy production comes from latent heat due to large scale and convective precipitation . In the bistable region, the range of $\dot{S}_{mat}$  is  $(10 , 19)$ $10^{-3}$W m$^{-2}$ K$^{-1}$   and $(34 , 62)$ $10^{-3}$W m$^{-2}$K$^{-1}$  for the SB and W respectively, therefore a factor of $3$ larger in the W regime with  respect  to the SB regime. This confirms that $\dot{S}_{mat}$  may be a better indicator than temperature for discriminating the SB and W states as already discussed in \cite{LucFr}. Again, there is a range of values of $\dot{S}_{mat}$  -- from $19$ to $34$ $10^{-3}$W m$^{-2}$ K$^{-1}$ -- which is not allowed by the system.


In the bistable region, the SB and W states are quantitatively very different with respect to their physical properties.  The two disjoint attractors can be thus thought to represent two different worlds, with completely different dynamical and thermodynamical properties. Therefore we treat them separately and then describe how the system makes a transition between them. For this reason in the following two sub-sections dynamical and thermodynamical properties of the manifolds will be analysed individually in terms of the vertical and horizontal surface temperature differences, Carnot efficiency, meridional heat transport and dissipation of kinetic energy. Furthermore we shall relate these properties to the average mean global temperature and the material entropy production. As is illustrated as solid and dashed purple lines, in the CS space figures, each manifold will be divided up in to two sub regions: W, W/Bistable, on the upper manifold and SB, SB/Bistable on the lower. Then in a third section, we will analyse the transitions between the two manifolds occurring in bifurcations regions.

\subsection{The warm state}

The meridional energy transport profile is worked out, as explained in \cite{Luc10}, by integrating over latitude the longitudinally averaged TOA radiation budget.  We then define as a scalar indicator of the transport half of the sum of the peak values of the poleward heat transport in the two hemispheres, $MET$.   In a moist atmosphere, the average global temperature and the meridional surface temperature difference, defined in our case as the difference between the mean surface temperature of the tropical (30S,30N) and the polar (90S,60S)  and (60N,90N) regions (see Figure \ref{quinto_a}),  are the main contributing factors for controlling the meridional heat transport (see Figure \ref{quarto_a}). This is due to the fact that temperature controls the latent heat released in the atmosphere because of the Clausius-Clapeyron effect \citep{HeldSoden} and the meridional temperature gradient controls   baroclinicity of the atmosphere \citep{Stone}. Additionally, another modulating factor is the vertical stratification of the atmosphere, as conditions of low stratification in the midlatitudes support stronger baroclinic activity for a given meridional temperature gradient \citep{Holton}. In the bistable region of the warm sector meridional heat transport has a flat response to increasing $S^*$ and [CO$_2$] and therefore $T_s$. With increased $T_s$, water vapour concentration of the atmosphere increases, thus leading to the strengthening of the poleward latent heat fluxes. In addition, the increased $T_s$ causes sea and continental ice as well as seasonal snow cover to retreat towards the poles, thus lowering the surface albedo gradient. This contributes negatively to the changes in the meridional transport, because it leads to a decrease in the baroclinicity. 

In the W regime, the boundary between the bistable and the monostable regime approximately marks the point at which the Earth surface loses its permanent sea-ice cover, thus supporting the idea that the presence of bistability is intimately linked with the powerful ice-albedo feedback. For $T_s$ larger than approximately $300$  K, the meridional temperature gradient decreases at a far slower rate with increasing $T_s$. This essentially means that in this region the meridional heat transport is controlled only by the availability of water vapour in the atmosphere. Therefore at temperatures above  $300$K, the meridional heat transport becomes decoupled from the surface meridional temperature difference. This analysis consequently shows how important the involvement of the hydrological cycle is in the magnitude of the meridional heat transport and moreover, it indicates that when going from warm to very warm climates the hydrological cycle becomes the dominant climatic feature, leading to strong positive dependence of the meridional heat transport on the surface temperature. Our results agree with the findings of \cite{Caballero05}, who found in aqua planet simulations, that there is little scope for reducing the meridional temperature gradient further once the sea-ice and snow have melted. For a constant meridional temperature gradient, an increase in the meridional latent heat fluxes with increasing global mean temperature was shown.

The conclusions drawn above find further support when looking at the mid-latitudes vertical temperature difference, $\Delta T_v$, defined as the mean temperature difference between the surface and the $500$ hPa level (see Figure \ref{quinto_b}) and which provides a rough measure of the stratification of the atmosphere of the atmosphere in the baroclinically active regime. The vertical temperature difference is largest along a band of the CS space centered half way through the bistable region in the direction of $S^*$. Our current climate conditions would appear to be positioned at the centre of the band peak where such temperature difference peaks, implying conditions of reduced vertical stratification. For  climates colder than present, increasing surface temperature causes the melting of sea-ice and of seasonal snow cover, so that the ensuing decrease in surface albedo (leading to increased surface absorption) accounts for the increasing vertical temperature difference. Instead, in climates colder than present, the decrease in equatorial vertical temperature difference with $T_s$ can be understood in terms of increased moist convection from warmer surface temperatures, resulting in an increase of moisture fluxes to the upper atmosphere, which then condense and release latent heat. 
In the bistable region, for the reasons discussed above, we expect to find a pronounced weakening of the dynamics of the climate system with increasing surface temperature. We test this hypothesis by computing the strength of the Lorenz energy cycle (Fig. \ref{quarto_b}), which is equal to the average rate of dissipation of kinetic energy, and the Carnot efficiency of the system (Fig. \ref{sesto_a}), which measures, instead, how far from equilibrium the system is. The dissipation, similarly to the meridional temperature gradient, decreases monotonically with the surface temperature, and reaches its largest value just at the cold boundary of the W manifold. The efficiency is maximized along a band of climate states characterised by slightly higher temperatures. This region in the CS space lies about half way between the peak in the meridional and vertical temperature gradients, since a large value of efficiency, by its very definition, requires a compromises between maximizing these two gradients. The efficiency decreases monotonically with increasing CO$_2$ or $S^*$: warmer climates are characterised by smaller temperature differences, since the transport of water vapour acts as a very efficient means for homogenising the temperature across the system. The system has lower ability to produce mechanical work and is characterised by very strong irreversible processes, as described by the very large values of entropy production realised in these conditions.  Consequently the value of the parameter of irreversibility $\alpha$  increases as conditions becomes warmer and warmer (not shown); see also \cite{LucFr}.

\subsection{The snowball state}

The SB state is intrinsically simpler than the W state because the hydrological cycle has a negligible influence. This is due to the fact that atmospheric temperatures are so low that we have in all cases an almost dry atmosphere.       As a result we have, e.g. that, as opposed to the W state, the parameter $\alpha$ is small ($\sim 1$) and weakly dependent on the value of the parameter $S^{*}$ and [CO$_2$] (not shown); see also \cite{LucFr}.

Moreover, in the SB state the meridional gradients (and not only the globally averaged values) of albedo are very low and depend weakly on $S^*$  and the CO$_2$, because under all conditions the sea surface is frozen almost everywhere and the continents are covered by ice and snow. The meridional heat transport (Figure \ref{settimo_a}) is much smaller than for the corresponding W states, because in the SB state a large fraction of the radiation is reflected back to space and the albedo gradients are small, so that the energy imbalance between low and high latitudes is small.
Throughout the SB state, increases in  $S^*$  and  [CO$_2$] , which lead to an increase in the surface temperature and in the rate of entropy production are accompanied by increases in the meridional heat transport and in the dissipation of kinetic energy (Figure \ref{settimo_b}). This tells us that with increased meridional heat transport the intensity of the circulation also increases. However we find that the meridional temperature difference (not shown) has a very weak dependence on changing [CO$_2$] and $S^*$, as it varies by only $4$ K across the explored parameter range, so that the changes in the baroclinicity of the system cannot be due to changes in the meridional heat transport and in the intensity of the Lorenz energy cycle. The lack of large variations in the meridional surface temperature profile are essentially due to the fact that  the net input of shortwave radiation is kept almost fixed by the constant surface albedo and by the virtue of almost no cloud cover. The system therefore can be seen as rather rigid, as changes in the absorbed radiation are almost exactly compensated by  changes in the meridional heat transport. The re-equilibration mechanism must then contain elements which are not present in the classic baroclinic adjustment \citep{Stone}. We find that the midlatitudes vertical temperature gradient (Figure \ref{ottavo_b}) increases substantially with increased value of  $S^*$  and [CO$_2$], as an effect of the increased absorption of radiation near the surface. The reduced vertical stratification leads to more pronounced baroclinic activity even for a fixed meridional temperature gradient, thus leading to an increase in the meridional heat transport and in the intensity of the Lorenz energy cycle. The argument is in this case quite straightforward because we are considering an almost  dry atmosphere. The efficiency has a dependence on $S^*$  and [CO$_2$] which, as in the W case, is intermediate between those of the meridional and vertical temperature gradient fields. In the case of the SB state, the main signature is given by the vertical gradient (Figure \ref{ottavo_b}).  This further reinforces the idea that the investigation of meridional temperature gradients is not enough to grasp the mechanisms through which the system generates available potential energy and material entropy production \citep{Luc10}.

\subsection{Transition and comparison between manifolds}

Up to now, the W and SB states have been characterised as two entirely distinct climate regimes, and have underlined that the basic mechanisms of re-equilibration are rather different. In this subsection we would like to present some ideas aimed at making sense of the transitions between the two states occurring when we get close to the boundary of the upper or of the lower manifold. As has been seen, the W$\rightarrow$SB transition is associated to a large decrease of surface temperature, rate of material entropy production, and meridional heat transport. This is intimately related to the fact that whereas in the W state the hydrological cycle is a major player of the climate dynamics, in the SB state the hydrological cycle is almost absent. Nonetheless, this does not say much about the processes leading to the transition between the two states, or better, describing how one of the attractors disappears. 

           We have also discovered that the usual dynamical indicators of the atmospheric state, i.e. the meridional temperature gradient and the vertical stratification do not necessarily indicate whether or not we are close to an irreversible transition of the system, e.g. by signaling something equivalent to a loss of ÒelasticityÓ of the system. In this regard, it is much more informative to observe how the efficiency behaves near the transitions. We find that, as a general rule, each transition is associated to a notable decrease (more than 30$\%$) of the efficiency of the system, and the closer the system gets to the transition in the CS space, the larger is the value of the efficiency. This can be interpreted as follows. If the system approaches a bifurcation point, its positive feedbacks become relatively stronger and the negative feedbacks, which act as re-equilibrating mechanisms, become less efficient. As a result, the differential heating driving the climate is damped less effectively, and the system is further from equilibrium, since larger temperature differences are present. Therefore, the system produces more work, thus featuring an enhanced Lorenz energy cycle and a stronger circulation. At the bifurcation point, the positive feedbacks prevail and the circulation, even if rather strong, is not able to cope with the destabilising processes, and the transition to the other manifold is realised. The new state is, by definition, more stable, and thus closer to equilibrium. The decreased value of the efficiency is exactly the marker of this property. This confirms what has been proposed by \cite{LucFr} but in much greater generality and can be conjectured to be a rather general properties of non-equilibrium systems featuring structural instabilities.

\section{Parameterisations}           
 \label{param}          
           
 In the previous sections, we have presented a systematic investigation of the thermodynamical properties of the planet in the SB and W states, obtained by exploring the CS parametric space. As mentioned above, in many cases  the main quantity controlling the thermodynamic properties is the average surface temperature $T_s$ only. This statement can be substantiated by looking at Figs. \ref{secondo}-\ref{ottavo} and observing that in most cases the isolines of the depicted thermodynamic quantities are approximately parallel, both in the W and SB state, to the isolines of $T_s$. This suggests the possibility of establishing approximate empirical laws of the form $\Gamma_{SB}(S^*,[CO_2])\approx \Gamma_{SB}(T_s(S^*,[CO_2]))$ and $\Gamma_W(S^*,[CO_2])\approx\Gamma_W(T_s(S^*,[CO_2]))$, where $\Gamma$ is a thermodynamical property such as, e.g. entropy production, and the lower index refers to whether we are in the SB or W state. For a given $\Gamma$, the empirical laws will be in general different in the SB vs in the W state. This result is quite interesting in a classical perspective of climate dynamics, where it is customary to parameterise large scale climate properties as a function of the surface temperature, especially when constructing simple yet meaningful models \citep{Saltzman}.

 We want to complement this approach by testing whether it is possible to construct empirical laws  expressing the thermodynamical properties of the system as a function of its emission temperature $T_E=\left(LW_{TOA}/\sigma  \right)^{1/4}$ -- where $LW_{TOA}$ is the area- and time-averaged outgoing long wave radiation at the top of the atmosphere and $\sigma $ the Stefan-Boltzmann constant -- , which is a quantity of more readily astrophysical interest and more fundamental physical significance. Therefore, we would like to be able to observe well-defined functions of the form $\Gamma_{SB}(S^*,[CO_2])\approx \Gamma_{SB}(T_E(S^*,[CO_2]))$ and $\Gamma_W(S^*,[CO_2])\approx \Gamma_W(T_E(S^*,[CO_2]))$, which, given the observation made above, requires also the existence of an approximate relationship $T_E\approx T_E(T_s)$.

Indeed, Fig. \ref{dieci} shows that it is possible to observe an empirical, monotonic relationship between $T_s$ and $T_E$ in a vast range of climate states, including both the W and SB states.  Obviously, the correspondence between the two variables is not perfect, so that for each value of $T_s$ it is possible to have climates featuring a range of values of $T_E$, and vice versa.  The spread associated with the ensemble points is of order $10$ K  and $5$ K for $T_E$, and $20$ K and $10$ K for $T_s$   in the W and SB states respectively. A range of values of $T_E$ ($230-240$ K) and $T_s$ ($240-260$ K) are not permitted by the system as a result of the ice-albedo instability. We can now proceed to see to which extent the quantities $T_s$ and  $T_E$ can be used as a good predictor of the thermodynamic properties of the planet and compare their skill  with $T_s$. The following description is, on purpose, purely qualitative, because we mostly want to show to what extent $T_E$ and  $T_s$ can be used as predictor, rather than discussing in detail what are the appropriate transfer functions.

Figure \ref{undici_a}  and \ref{undici_b} show  that a clear monotonic relationship exists between $T_E$, $T_s$ and the material entropy production $\overline{\dot{S}_{mat}}$, with higher values of $T_E$ and $T_s$ associated with larger values of $\overline{\dot{S}_{mat}}$.    It is important to note that while the values of $\overline{\dot{S}_{mat}}$ and $T_s$ are related to a very high degree of approximation on a one-to-one basis, the functional dependence between $T_E$ and  $\overline{\dot{S}_{mat}}$ is less well defined (the spread  of points is much wider than in the $T_s$ case, and especially so in the very warm regimes).     Note also that the dependences,  while monotonic in both cases, are very different between the W and SB states. The common line is that the more is the energy absorbed by the planet (larger values of $T_E$), the more intense in the production of material entropy.              Note that this is indeed not the case when considering the intensity of the Lorenz energy cycle (Fig. \ref{dodici_a} and \ref{dodici_b}): higher absorption of energy corresponds to a weaker energy cycle in the W state and to a stronger cycle in the SB state. Therefore, the increase in the entropy production with $T_E$ is due, in the SB case, to the increased dissipation of kinetic energy, while in the W state it is due to the greatly enhanced hydrological cycle. In agreement with what has been discussed in the previous section, one can explain the contrasting responses of the W and SB states to increases in the absorption of energy by analyzing the Carnot efficiency of the planetary fluid (Fig. \ref{tredici}). 
It is interesting to note that, at the transitions,  the values of the efficiency are practically equal:   $\eta_{W\rightarrow SB}\approx \eta_{SB,W}\approx 0.03$. Moreover, we observe that $\eta$ saturates in the very cold regime of SB states and very warm regime of W (see Fig.~\ref{tredici})  at $\eta_{sat,W}\approx \eta_{sat,SB}\approx 0.012$. Also for the Lorenz energy cycle and the efficiency, $T_s$ does a much better job as predictor of the thermodynamic quantities, since the spread of points is much higher when considering $T_E$.

Last, we want to present the link between $T_E$, $T_s$ and the meridional energy transport $MET$ (Fig. \ref{quattordici}). We observe that in the SB regime there is a weakly positive relationship between $T_E$ ($T_s$) and the transport, for the reasons described in the previous section; instead, for a vast range of values of $T_E$ ($T_s$) in the W regime, the transport is almost insensitive to $T_E$ ($T_s$), as discussed thoroughly in \cite{Caballero05} and \cite{Battisti1,Battisti2}. The rigidity of the climate system has been attributed by \cite{Battisti1,Battisti2} to   the fact that the meridional heat transport is mainly determined by planetary albedo and thus atmospheric composition rather that surface albedo and sea-ice coverage.      However, such a mechanism ceases to exist when the ice albedo feedback becomes ineffective because of the disappearance of sea ice above a threshold value of $T_E\approx 255$ K ($T_s\approx 300$ K). Above this value, we observe a steep monotonic increase of the transport with temperature, because  changes in the latent heat transport are mainly responsible for this behavior. This agrees with the idea that the dynamics and sensitivity of a warm planet is, in some sense, dominated by the hydrological cycle. In this latter case, the skill of the two temperature quantities in parameterising the $MHT$ is comparable.

Figs \ref{undici}-\ref{quattordici} clearly define the transfer functions allowing to compute the thermodynamic properties of the planet from the value of $T_E$ or $T_s$ only: this  is definitely intriguing because it paves the way for computing non-equilibrium properties from a quantity describing the zero-order radiation balance of the system at the top of the atmosphere and of the radiative-convective equilibrium at surface (in the case of $T_s$).   Let us finally note that, in general, the functional dependences of the thermodynamic quantities  on $T_s$ is more precisely defined than   the one on $T_E$ since these  show a larger spread of the ensemble points.    This is particularly evident in quantities such as $\overline{\dot{S}_{mat}}$ (Fig.~\ref{undici}) and $\eta$ (Fig.~\ref{tredici}), $\overline{W}$ (Fig.~\ref{dodici})  but less in others, e.g. $MET$ (Fig.~\ref{quattordici}).  Figures (\ref{dieci}-\ref{tredici}) leave various open questions which definitely need to be accurately addressed in future studies: i) why is it possible to use to a good degree of approximation the quantity $T_E$ to parameterise the non-equilibrium thermodynamical properties of a system? ii) why does the surface temperature does an even better job (actually, astounding) in controlling such complex properties? iii) is the result on the decrease of the efficiency at the bifurcation points in both W$\rightarrow$SB and SB$\rightarrow$W transitions of more general relevance?

\section{Summary and conclusions}
\label{concl}

Motivated by  the recent discoveries of extrasolar planets, in this contribution we have shown that the habitability conditions,  for an Earth-like planet, is a necessary but not solely sufficient condition  since two stable sets of states are possible, with one of the two characterized by the global glaciation of water.    In particular we have shown how the stability properties of the climate system depend on the modulation of the two main parameters describing the radiative forcing, i.e. the energy input of the parent star and the long wave opacity of the atmosphere. In our analysis we propose that the point of view of non-equilibrium thermodynamics is especially useful for understanding the global properties of the climate system and for interpreting its global instabilities.

              We have discovered that in a rather wide parametric space, which includes the present climate conditions, the climate is multistable, i.e. there are two coexisting attractors, one characterised by warm conditions, where the presence of sea ice and seasonal snow cover is limited (W state), and one characterised by a completely frozen sea surface, the so-called snowball (SB) state. Our results, obtained using the climate model PlaSim, confirm and extend what has been obtained in various studies performed with models of various degrees of complexity. We point the reader to \cite{Pierre11} and \cite{LucFr} for an extensive discussion.  In this regard, the main improvement of this work is that a two dimensional parametric space is explored (whereas usually variations in the solar constant or in the opacity of the atmosphere are considered separately), which allows the gathering of  more complete information on the possible states of the climate system and actual mechanisms relevant  in a paleoclimatological context, as explained in \cite{Pierre11}.
                 For all considered values of [CO$_2$], which range from $90$  to about $3000$ ppm, the width of the bistable region is about $200$ Wm$^{-2}$ in terms of the value of the solar constant, and its position depends linearly on the logarithm of the [CO$_2$], being centered around smaller values of the solar constant for increasing opacity of the atmosphere, shifting by about $15$ Wm$^{-2}$ per doubling of CO$_2$ concentration. The W state is characterized by surface temperature being $40$ K-$60$ K higher than in the SB state, and also, in terms of the material entropy production, is larger by a factor of 4 (order of $40-60$ $10^{-3}$Wm$^{-2}$  vs.  $10-15$ $10^{-3}$Wm$^{-2}$). The boundaries of the bistable region are approximately isolines of the globally averaged surface temperature, and in particular, the warm boundary, beyond which the SB state cannot be realized, is characterized by vanishing permanent sea ice cover in the W regime. This reinforces the idea that the ice-albedo feedback is the dominant mechanism for the multistability properties.
                 
              The thermodynamical and dynamical properties of the two states are largely different, as if we were discussing two entirely different planets. In the W state the climate is dominated by the hydrological cycle and latent heat fluxes are prominent in terms of redistributing the energy in the system and as contributors to the material entropy production. The SB state is eminently a dry climate, where heat transport is realized through sensible heat fluxes and entropy is mostly generated through the dissipation of the kinetic energy. The dryness of the SB atmospheres also explains why the climate sensitivity is much smaller.
              
               In the bistability region of the W states, the meridional heat transport is rather constant, as the contrasting effect of the enhancement of latent heat fluxes driven by increasing surface temperature and the reduction in the baroclinicity due to the decrease in the meridional temperature gradient compensate almost exactly. In the W state beyond the bistability region, the meridional heat flux increases quite dramatically  with the surface temperature (Fig. \ref{quinto_a} and \ref{quattordici}) since  the compensating albedo mechanisms are shut off as sea ice is completely removed from the surface. In the SB states, increased incoming radiation or increased [CO$_2$]  leads to increases in the meridional heat transport. In this case, the water vapor plays little role, and, somewhat surprisingly, the meridional temperature difference has also a rather flat response to the parametric modulation. In this case, the dominant mechanism determining the properties of the meridional heat transfer is the change in the vertical stratification, which becomes weaker for warmer climate conditions. This implies that the atmospheric circulation strengthens for increasing values of the solar constant and of  [CO$_2$]. In fact,  the Lorenz energy cycle becomes stronger for warmer climate conditions, and the Carnot-like efficiency of the climate system has an analogous behavior. The opposite holds for the W state, where the intensity of the Lorenz energy cycle and the efficiency decrease for warmer conditions, the reason being that the water vapor becomes more and more efficient in homogenizing the system and destroying its ability to generate available potential energy. 
               
               A general property which has been  found is that, in both regimes, the efficiency increases when we get closer to the bifurcation point and at the bifurcation point the transition to the newly realized stationary state is accompanied by a  large decrease in the efficiency. This can be framed in a rather general thermodynamical context: the efficiency gives a measure of how far from equilibrium the system is. The negative feedbacks tend to counteract the differential heating due to the sun's insolation pattern, thus leading the system closer to equilibrium. At the bifurcation point, the negative feedbacks are overcame by the positive feedbacks, so that the system makes a global transition to a new state, where, in turn, the negative feedbacks are more efficient in stabilizing the system.


The results discussed in this paper support the adoption of new diagnostic tools based on non-equilibrium thermodynamics  for  analysing the fundamental properties of planetary atmospheres. The next step in this direction is a more quantitative understanding of the global relationships between surface temperature, material entropy production, meridional heat fluxes and Carnot-like efficiency  for the SB and W states and to propose possible parametrisations for the dynamical and thermodynamical properties of the  SB and W states.  Another line of research will explore  the dependence  of these quantities  on other fundamental parameters, e.g. the rotation rate and the surface drag, relevant for atmospheres of terrestrial planets.  Recent theoretical results suggest that it is possible to derive up to a good degree of approximation  the full thermodynamical  properties of  planetary atmospheres from the  coarse  observations of radiative fluxes at TOA and surface \citep{LucFr}. In this work, we have pursued this line of research by proposing well-defined empirical functions allowing for expressing the global non-equilibrium thermodynamical properties of the system in terms of the globally averaged surface temperature and planetary emission temperature only.   The skills of such temperature quantities as predictors is fairly good. The reasons why these parameterisations work so well, and why this is especially the case for  globally averaged surface temperature definitely require further investigations.     These findings fully confirm the results discussed in \cite{LucFr,GenSens} and generalize them to a much larger variety of climates, explored by changing both the energy input of parent star and the opacity of the atmosphere. These results reinforce the relevance of the globally averaged surface temperature as a meaningful thermodynamical quantity of the climate system, beyond its obvious practical importance. Moreover, they pave the way for the possibility of practically deducing fundamental properties of planets in the habitable zone from a relatively simple observable.

\subsection*{Acknowledgements}
This work was supported by the EU-FP7  ERC grant NAMASTE. The authors would like to thank  E. T. Cartman, K. Fraedrich, P. Hauschildt, F. Lunkeit  and F. Ragone for their comments and insightful discussions.

\clearpage

\section*{Figures' captions}
\begin{figure}
\centering
\subfigure[]{
 \includegraphics[angle=-0, width=0.9\textwidth]{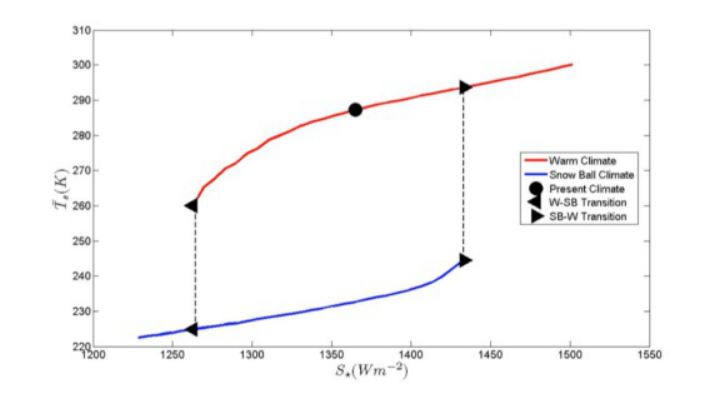}
  \label{primo_a}
   }
   \subfigure[]{
    \includegraphics[angle=-0, width=0.9\textwidth]{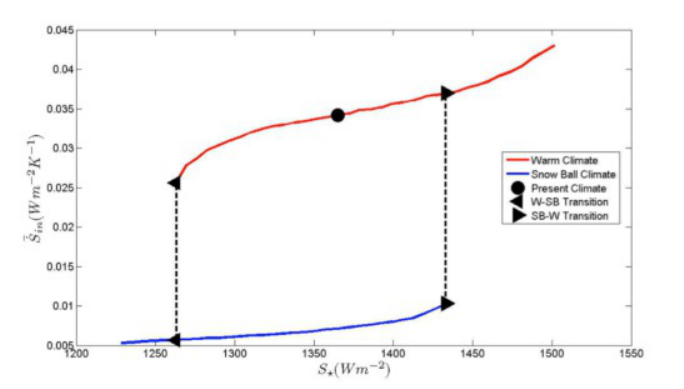}
    \label{primo_b}
     } 
\caption{  
  \label{primo}}
\end{figure}

\begin{figure}
\centering
\subfigure[]{
 \includegraphics[angle=-0, width=0.7\textwidth]{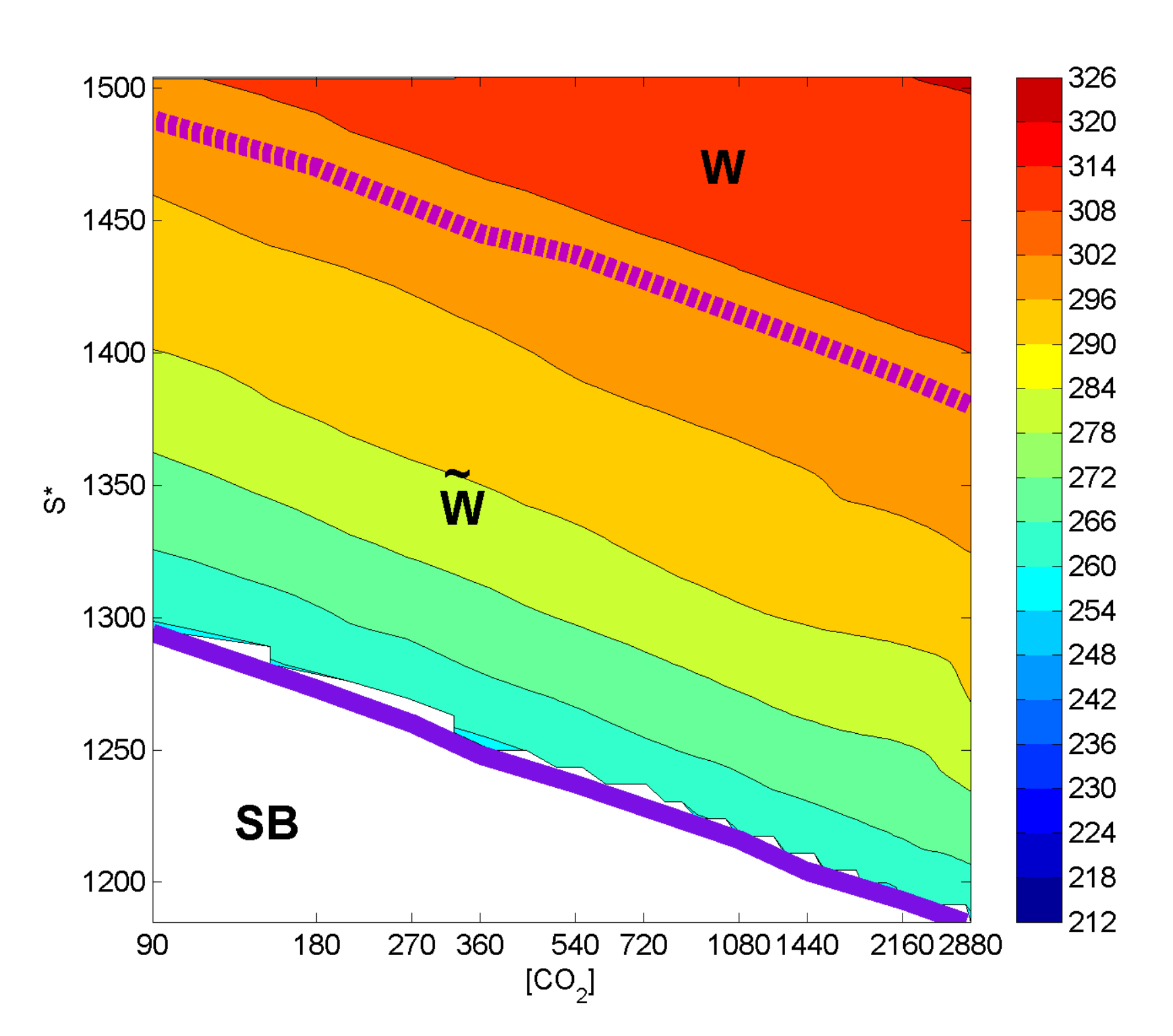}
  \label{secondo_a}
   }
   \subfigure[]{
    \includegraphics[angle=-0, width=0.7\textwidth]{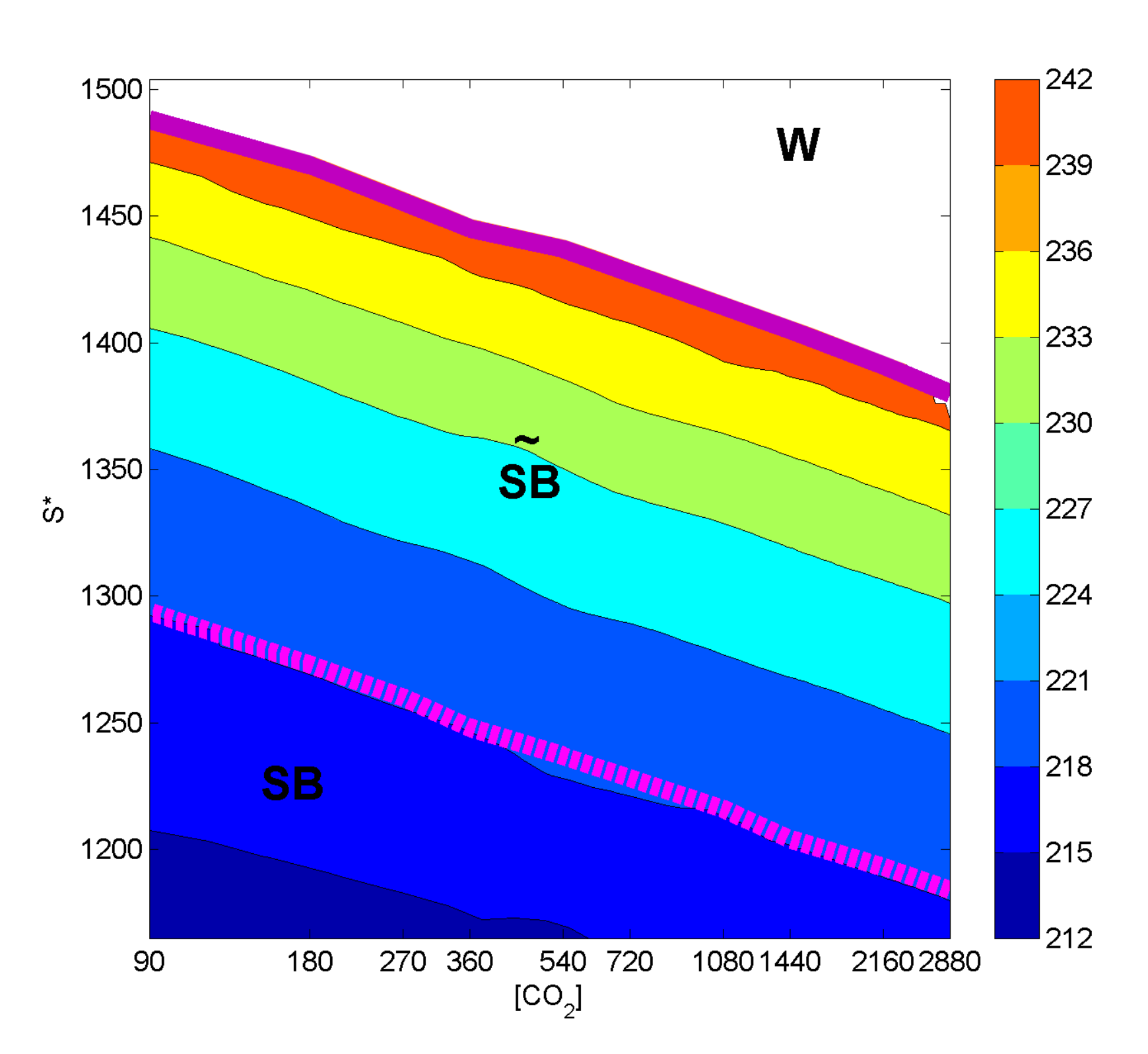}
    \label{secondo_b}
     } 
\caption{    
               \label{secondo}}
\end{figure}

\begin{figure}
\centering
\subfigure[]{
 \includegraphics[angle=-0, width=0.7\textwidth]{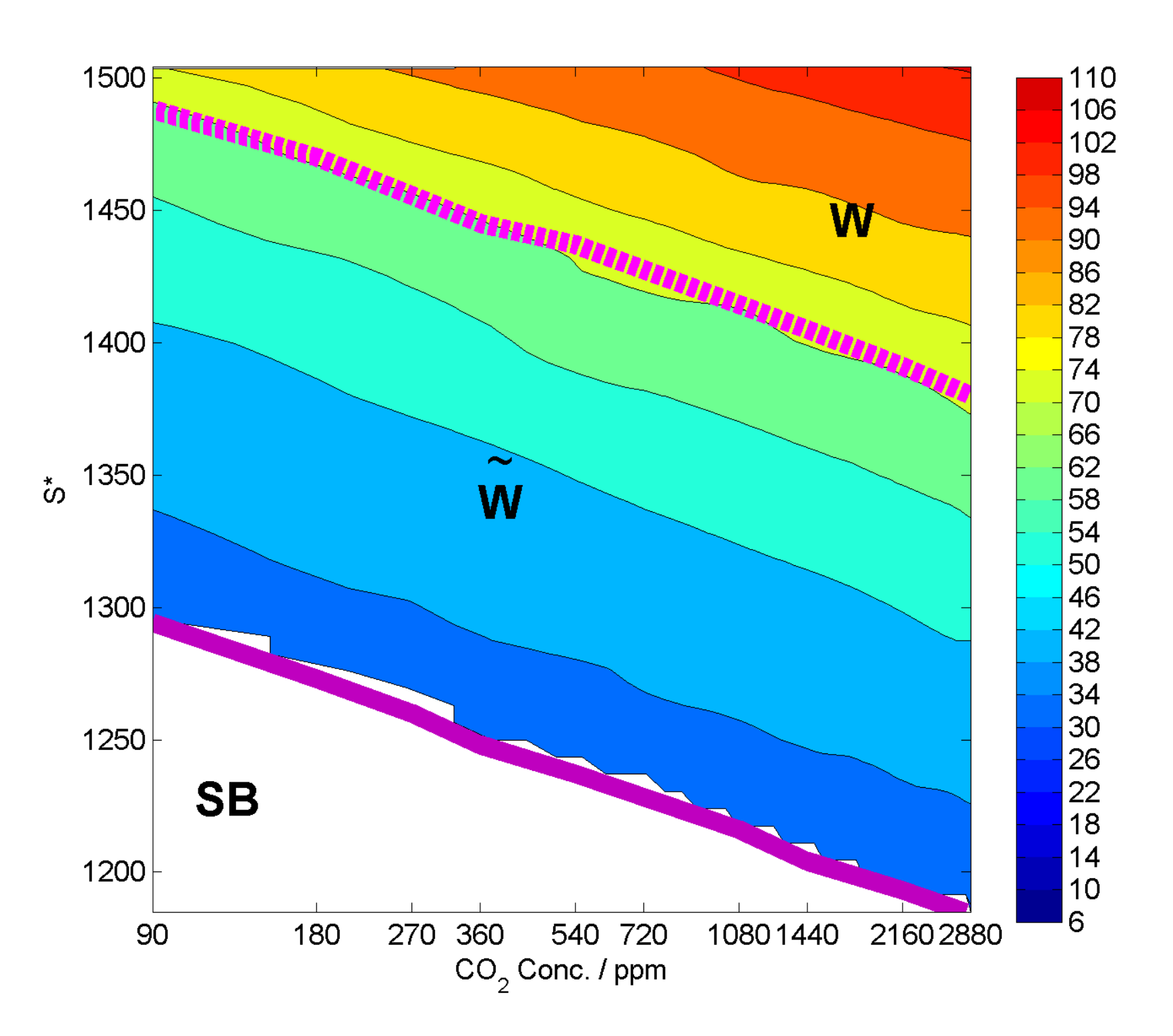}
  \label{terzo_a}
   }
   \subfigure[]{
    \includegraphics[angle=-0, width=0.7\textwidth]{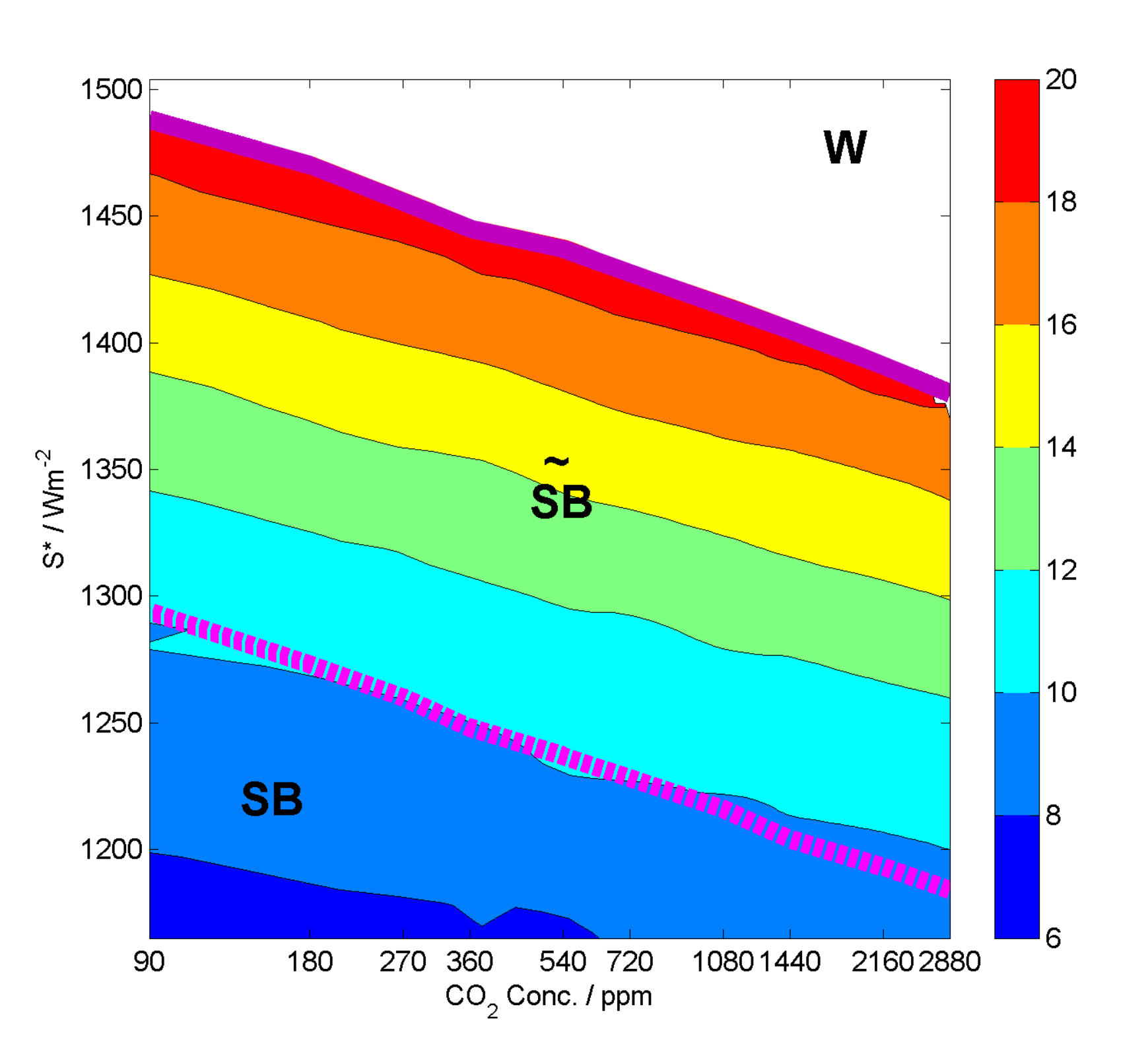}
    \label{terzo_b}
     } 
\caption{  
  \label{terzo}}
\end{figure}

\begin{figure}
\centering
\subfigure[]{
 \includegraphics[angle=-0, width=0.7\textwidth]{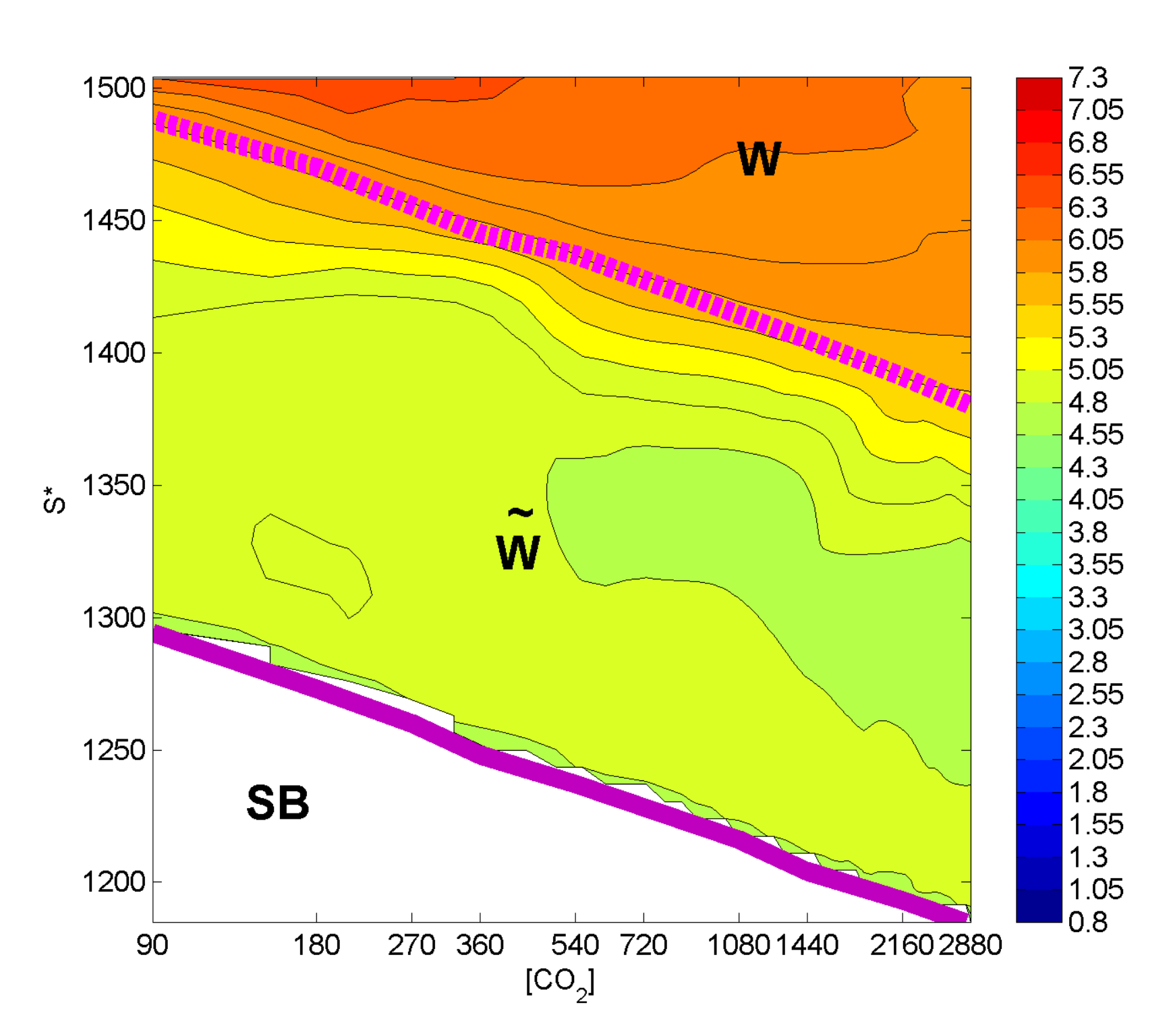}
  \label{quarto_a}
   }
   \subfigure[]{
    \includegraphics[angle=-0, width=0.7\textwidth]{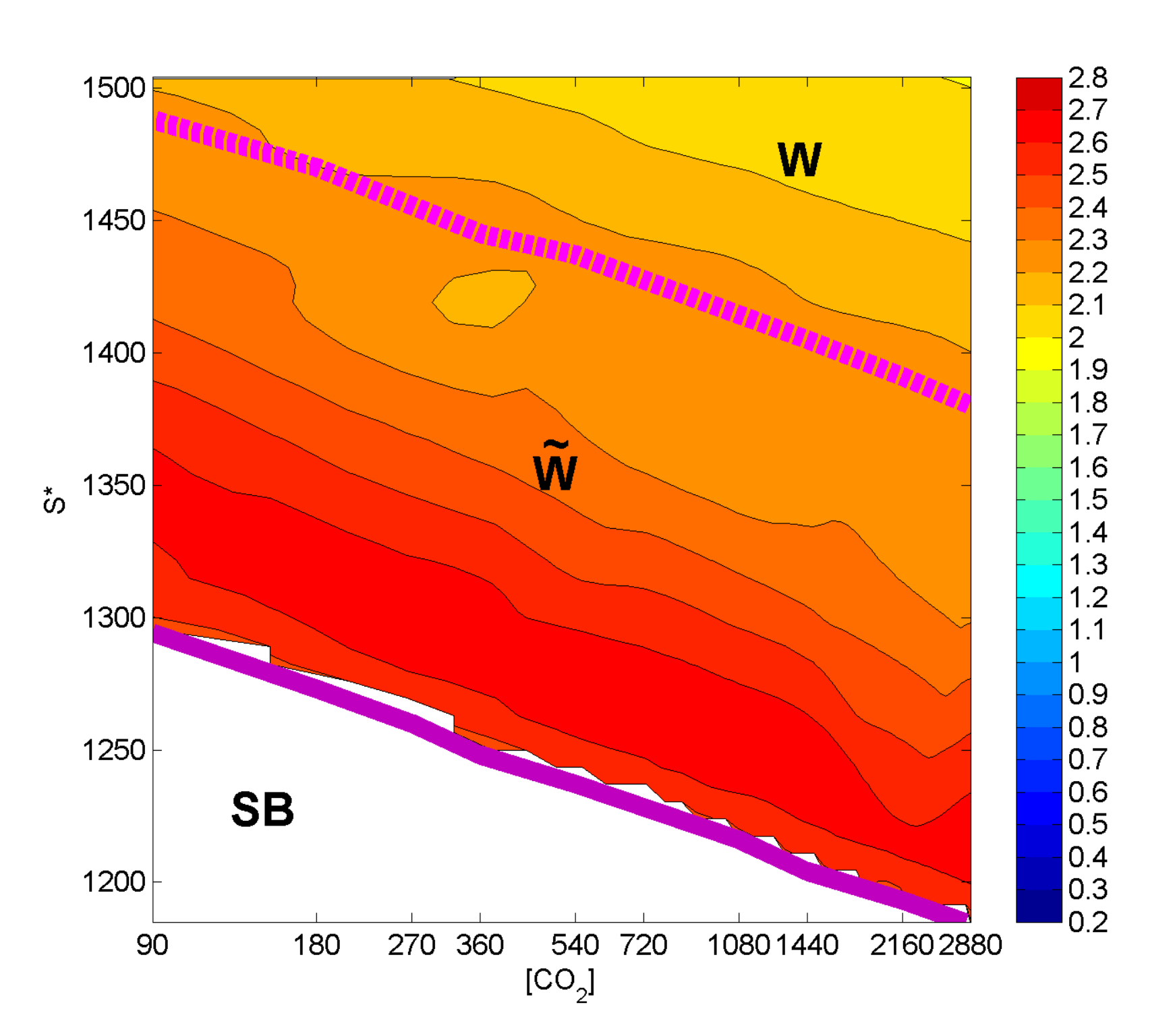}
    \label{quarto_b}
     } 
\caption{    
 \label{quarto}}
\end{figure}

\begin{figure}
\centering
\subfigure[]{
 \includegraphics[angle=-0, width=0.7\textwidth]{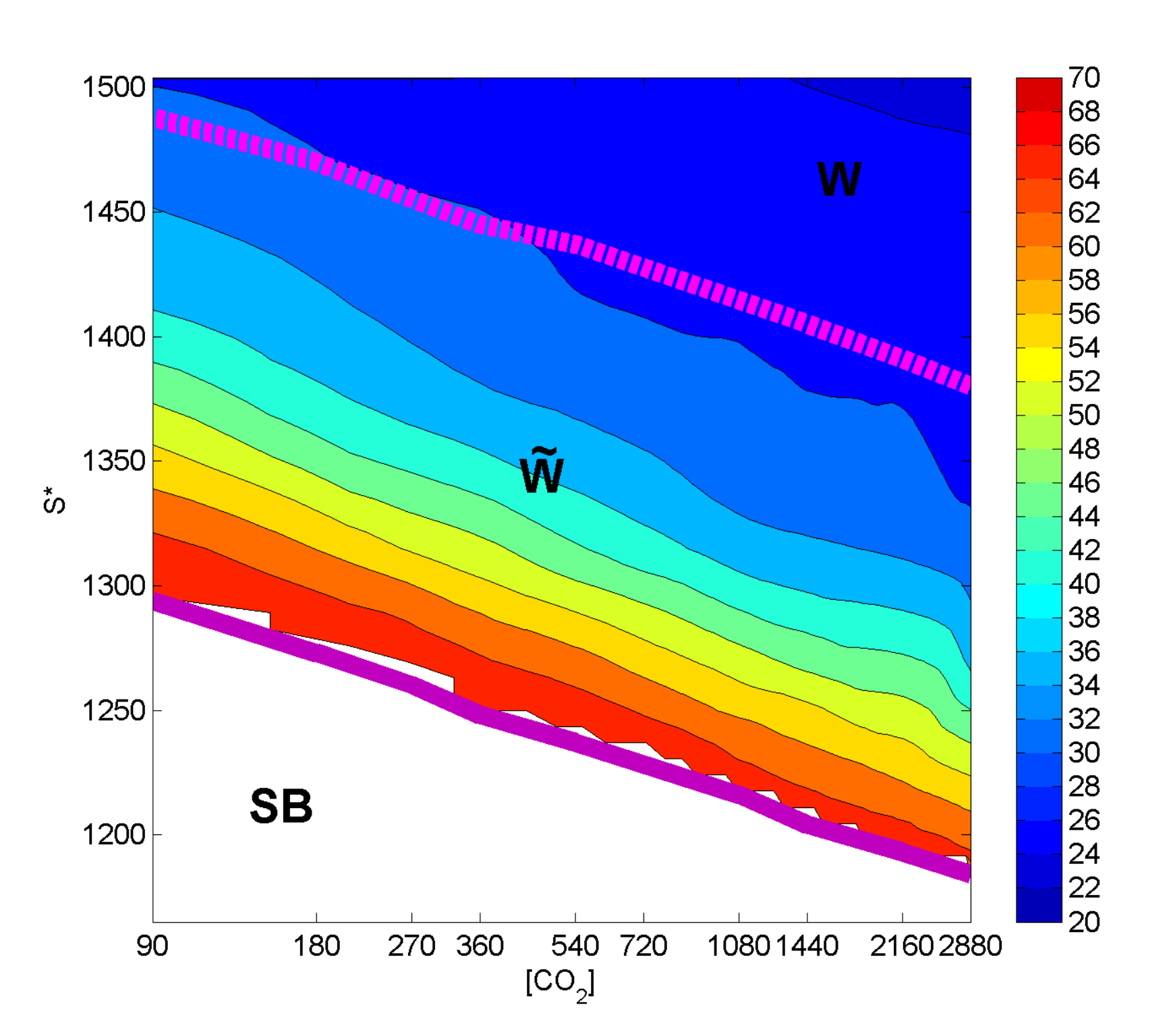}
  \label{quinto_a}
   }
   \subfigure[]{
    \includegraphics[angle=-0, width=0.7\textwidth]{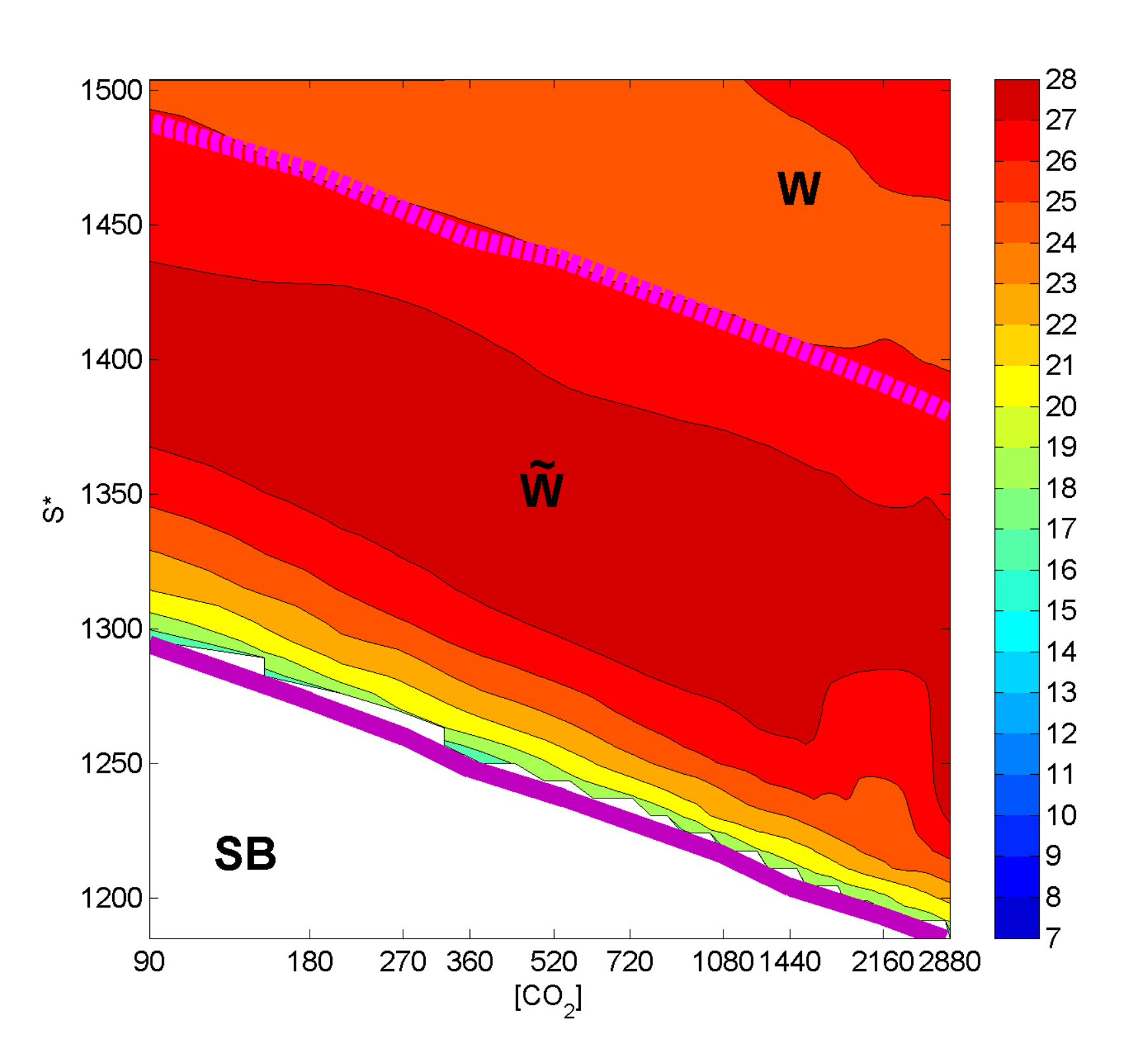}
    \label{quinto_b}
     } 
\caption{  
 \label{quinto}}
\end{figure}

\begin{figure}
\centering
\subfigure[]{
 \includegraphics[angle=-0, width=0.9\textwidth]{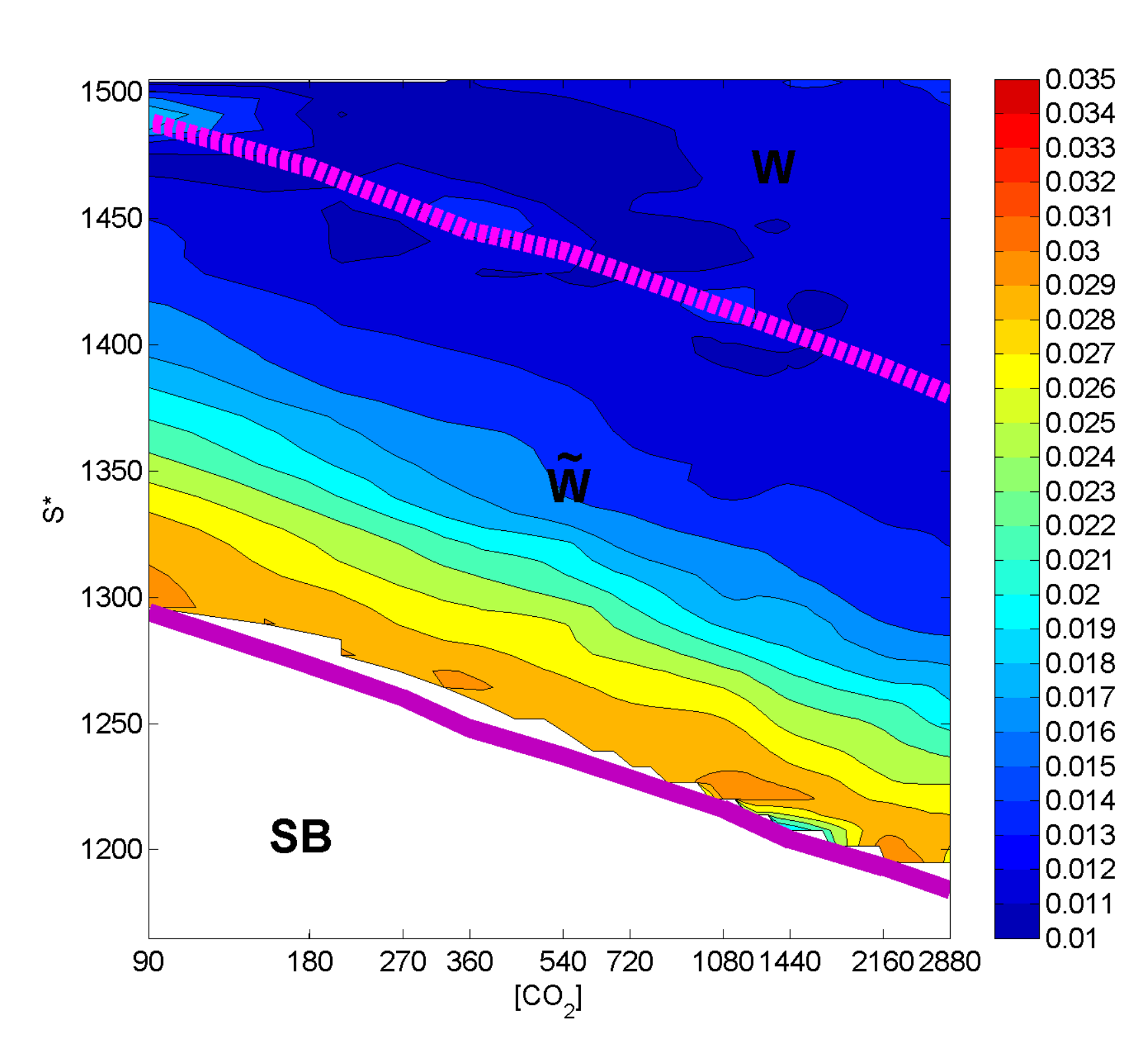}
  \label{sesto_a}
   }
\caption{   
 \label{sesto}}
\end{figure}

\begin{figure}
\centering
\subfigure[]{
 \includegraphics[angle=-0, width=0.7\textwidth]{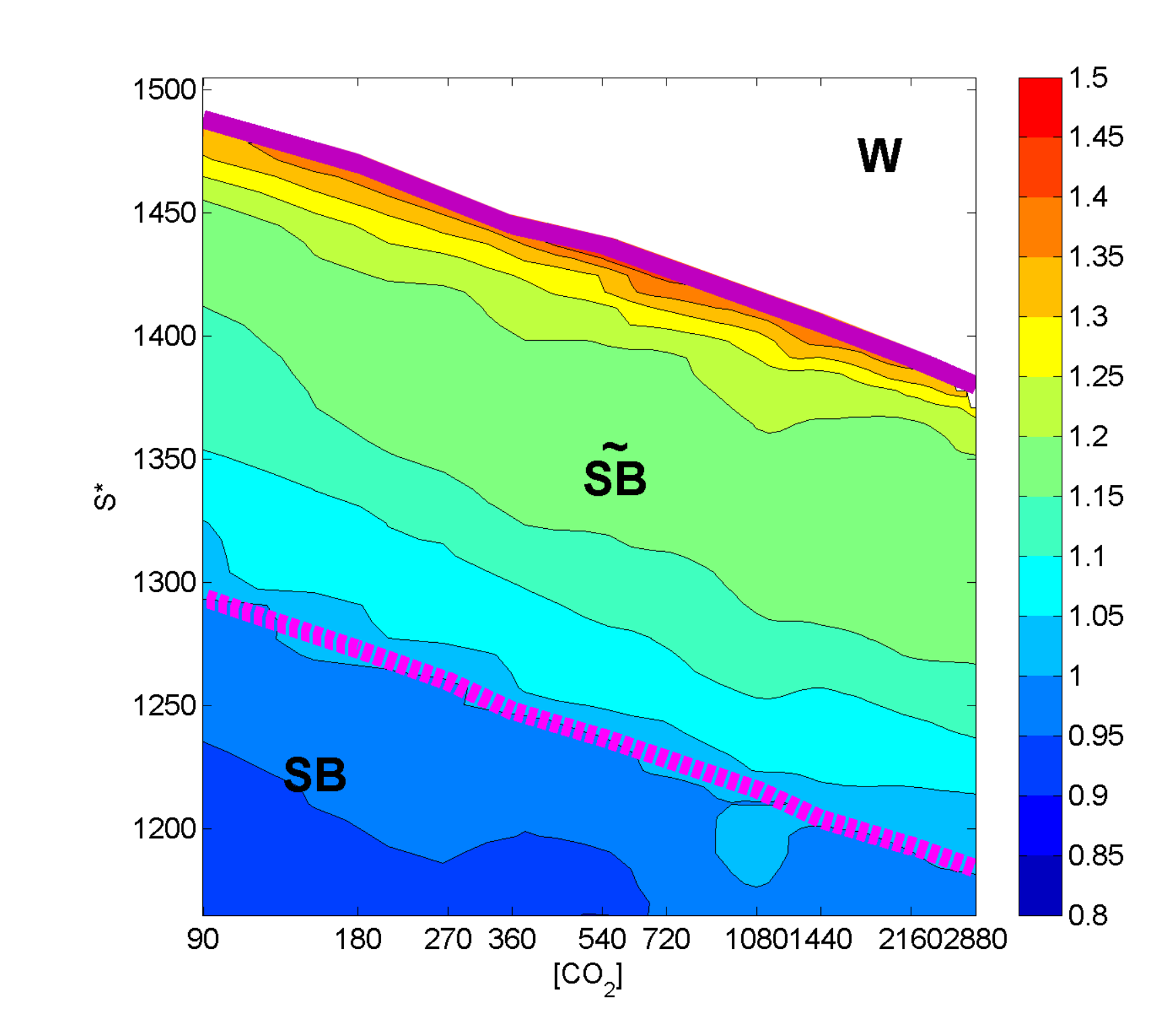}
  \label{settimo_a}
   }
   \subfigure[]{
    \includegraphics[angle=-0, width=0.7\textwidth]{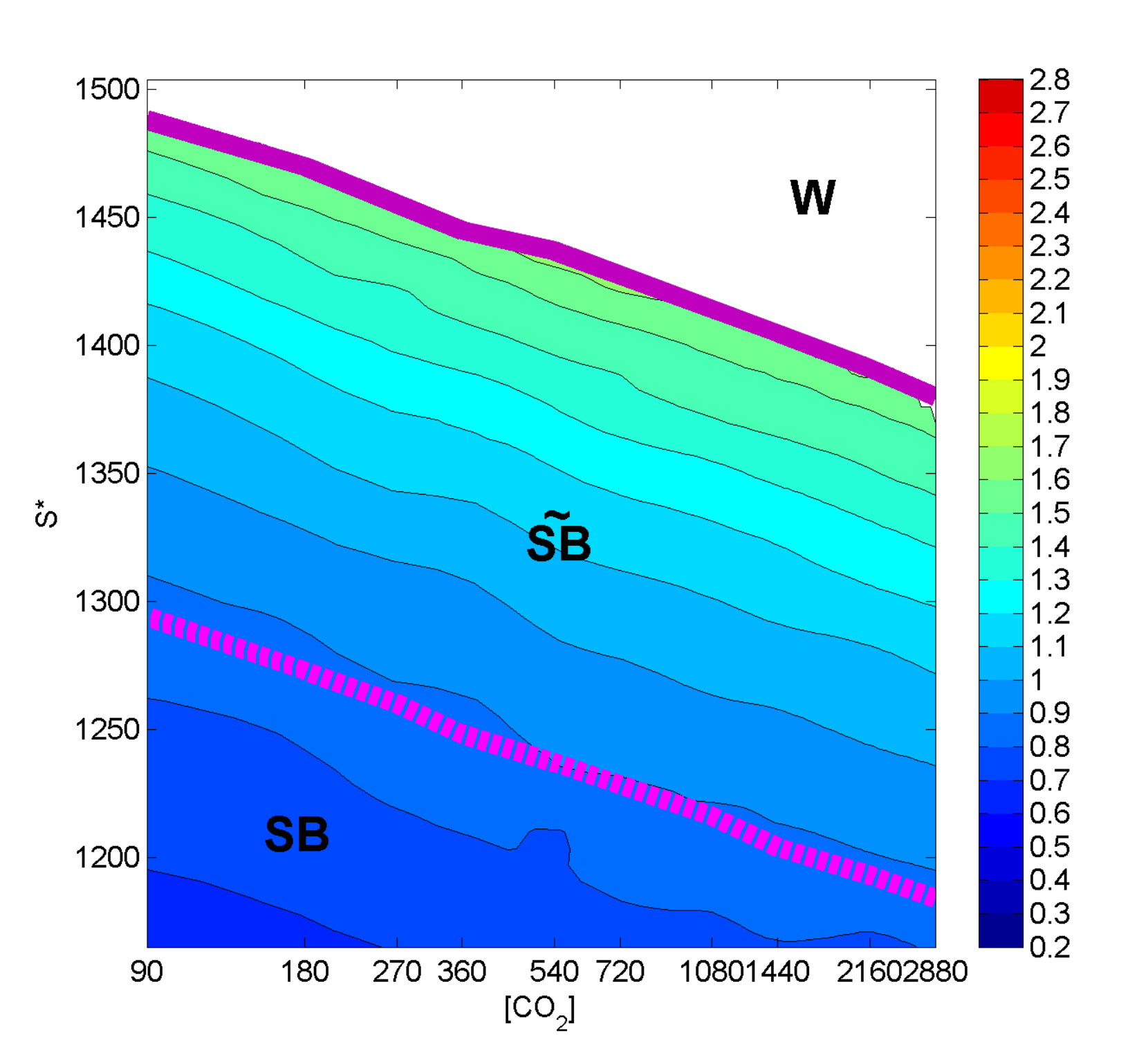}
    \label{settimo_b}
     } 
\caption{   
 \label{settimo}}
\end{figure}

\begin{figure}
\centering
\subfigure[]{
 \includegraphics[angle=-0, width=0.7\textwidth]{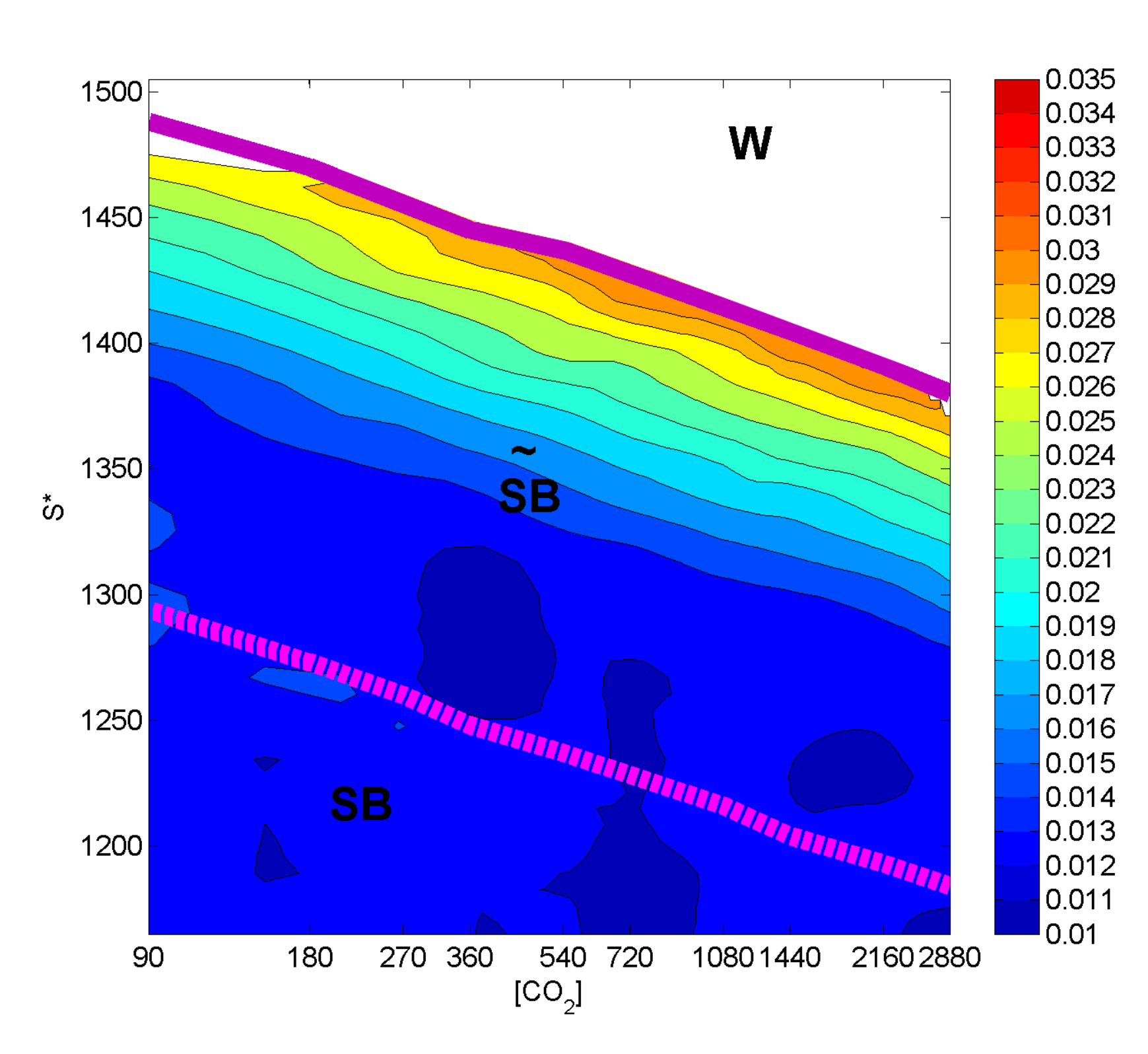}
  \label{ottavo_a}
   }
   \subfigure[]{
    \includegraphics[angle=-0, width=0.7\textwidth]{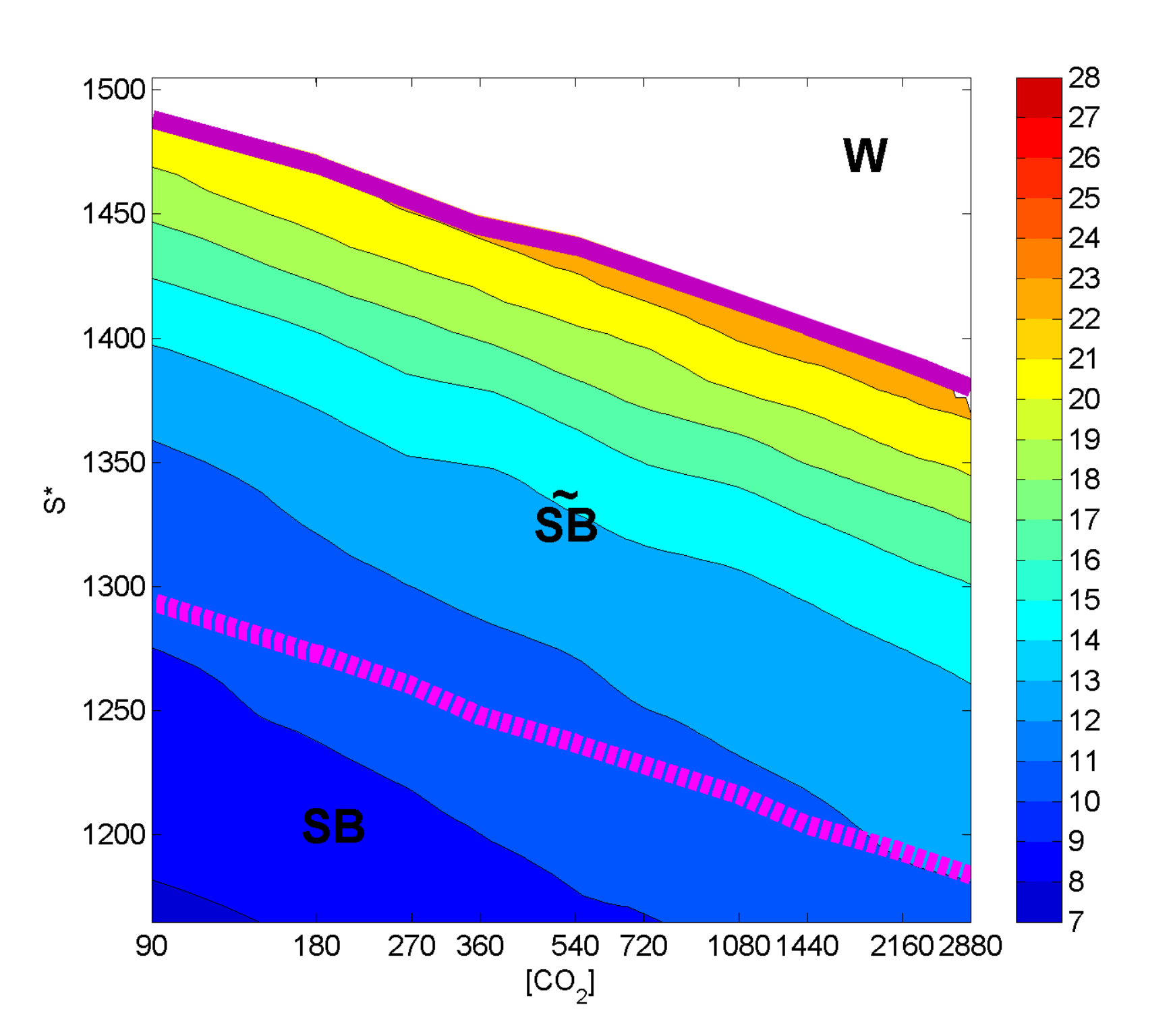}
    \label{ottavo_b}
     } 
\caption{  
 \label{ottavo}}
\end{figure}

\begin{figure}
\centering
 \includegraphics[angle=-0, width=0.9\textwidth]{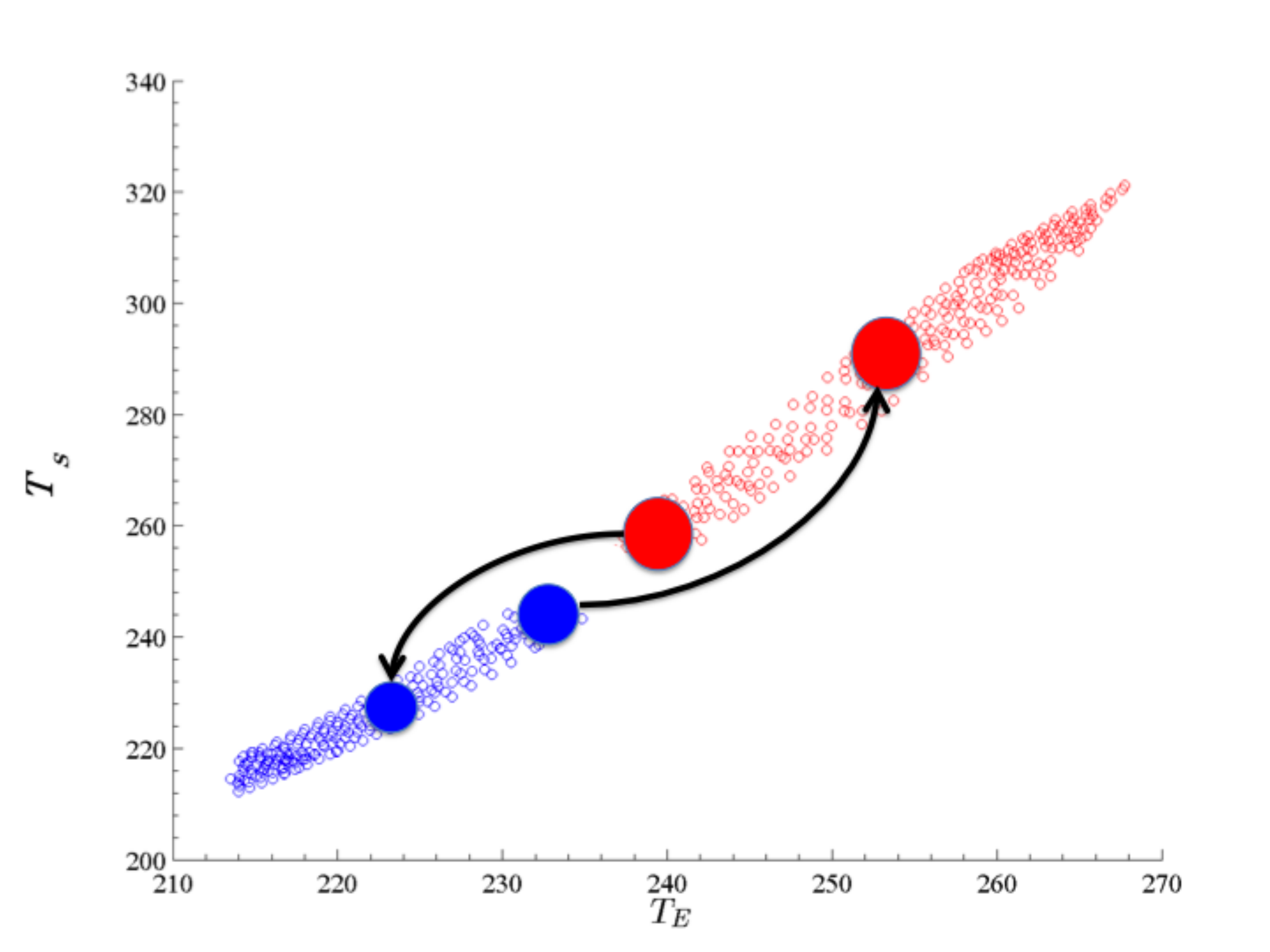}
 
\caption{   
 \label{dieci}}
\end{figure}

\begin{figure}
\centering
\subfigure[]{
 \includegraphics[angle=-0, width=0.7\textwidth]{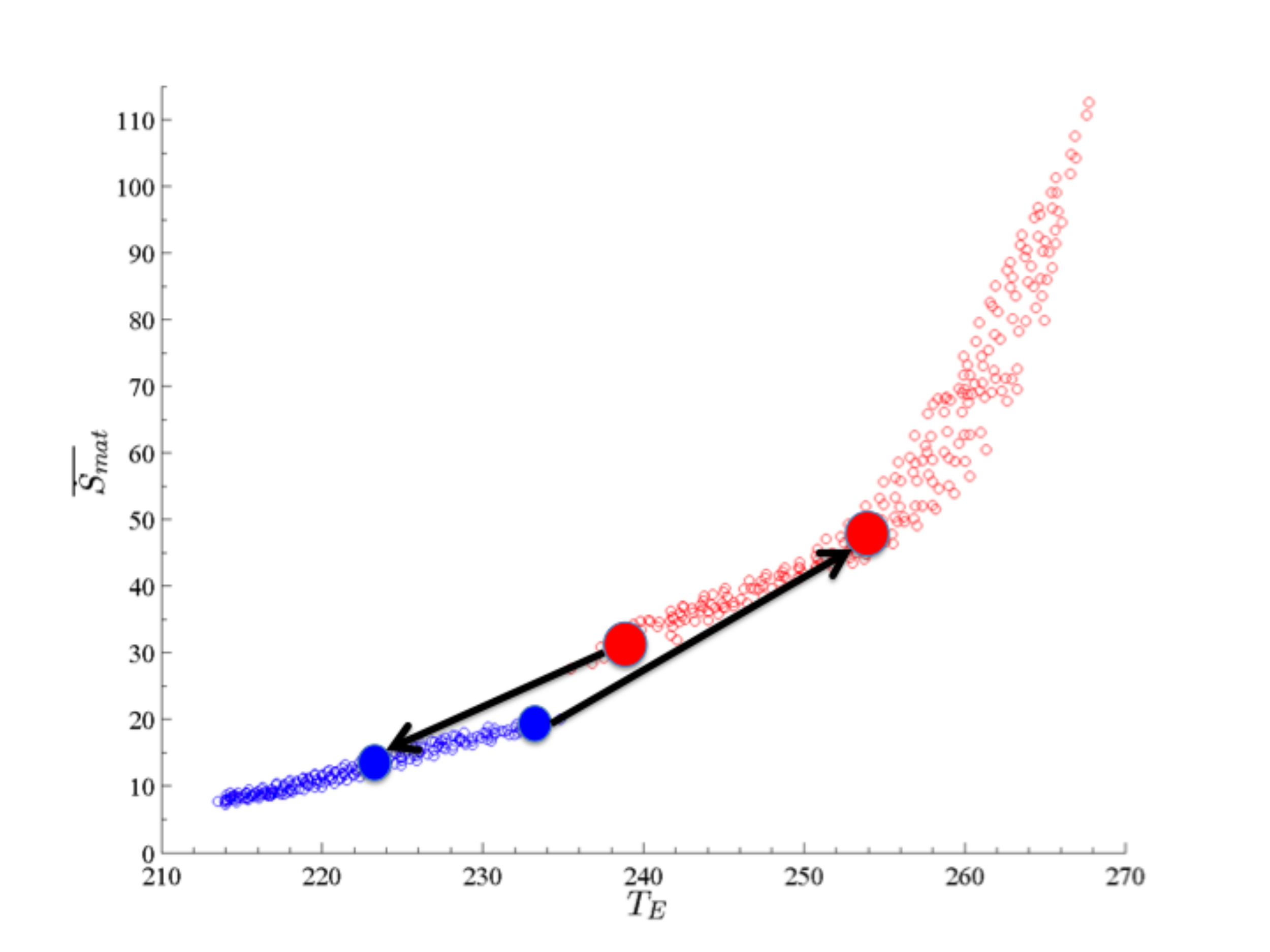}
  \label{undici_a}
   }
   \subfigure[]{
    \includegraphics[angle=-0, width=0.7\textwidth]{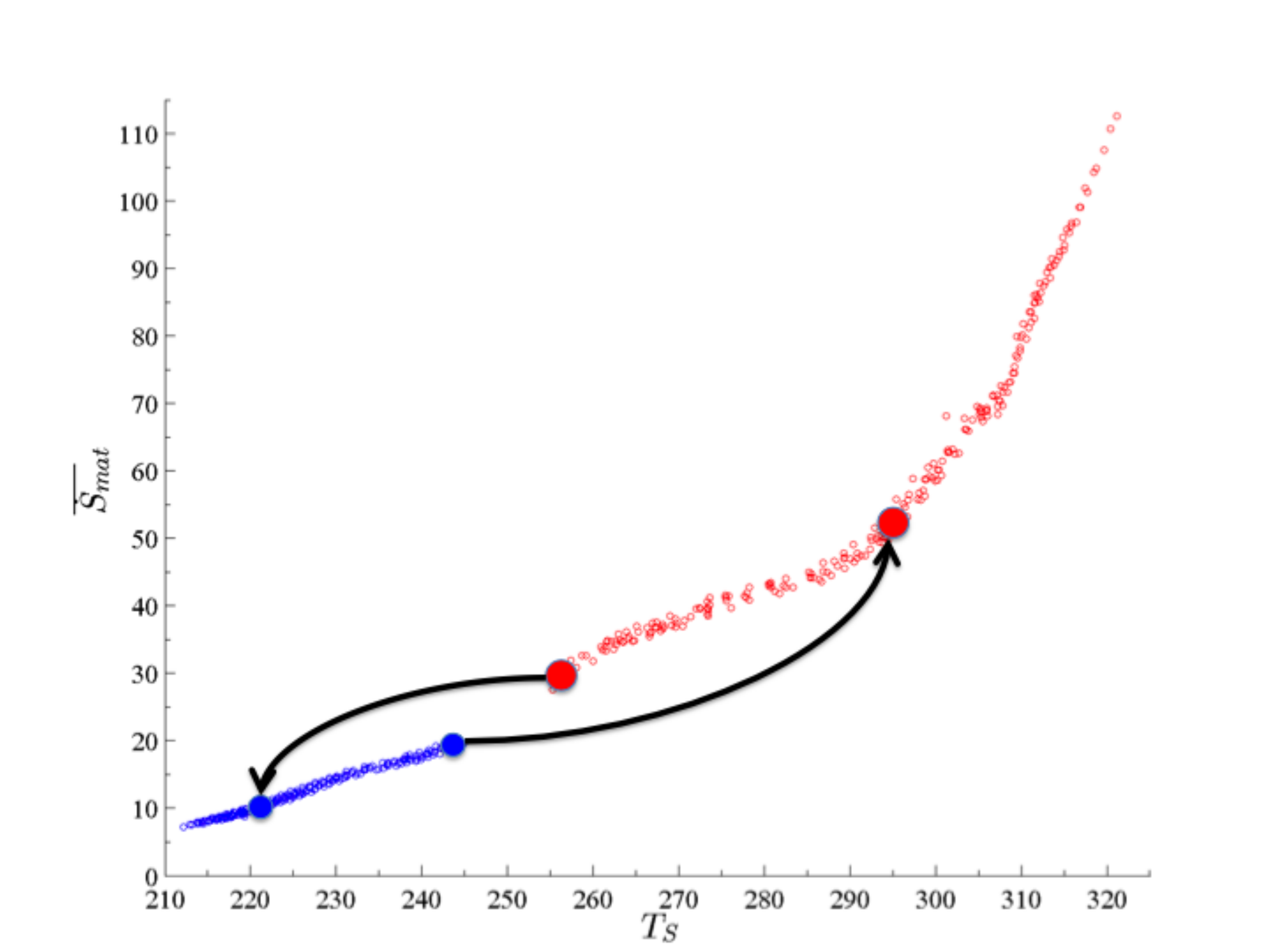}
    \label{undici_b}
     } 
\caption{ 
 \label{undici}}
\end{figure}

\begin{figure}
\centering
\subfigure[]{
 \includegraphics[angle=-0, width=0.7\textwidth]{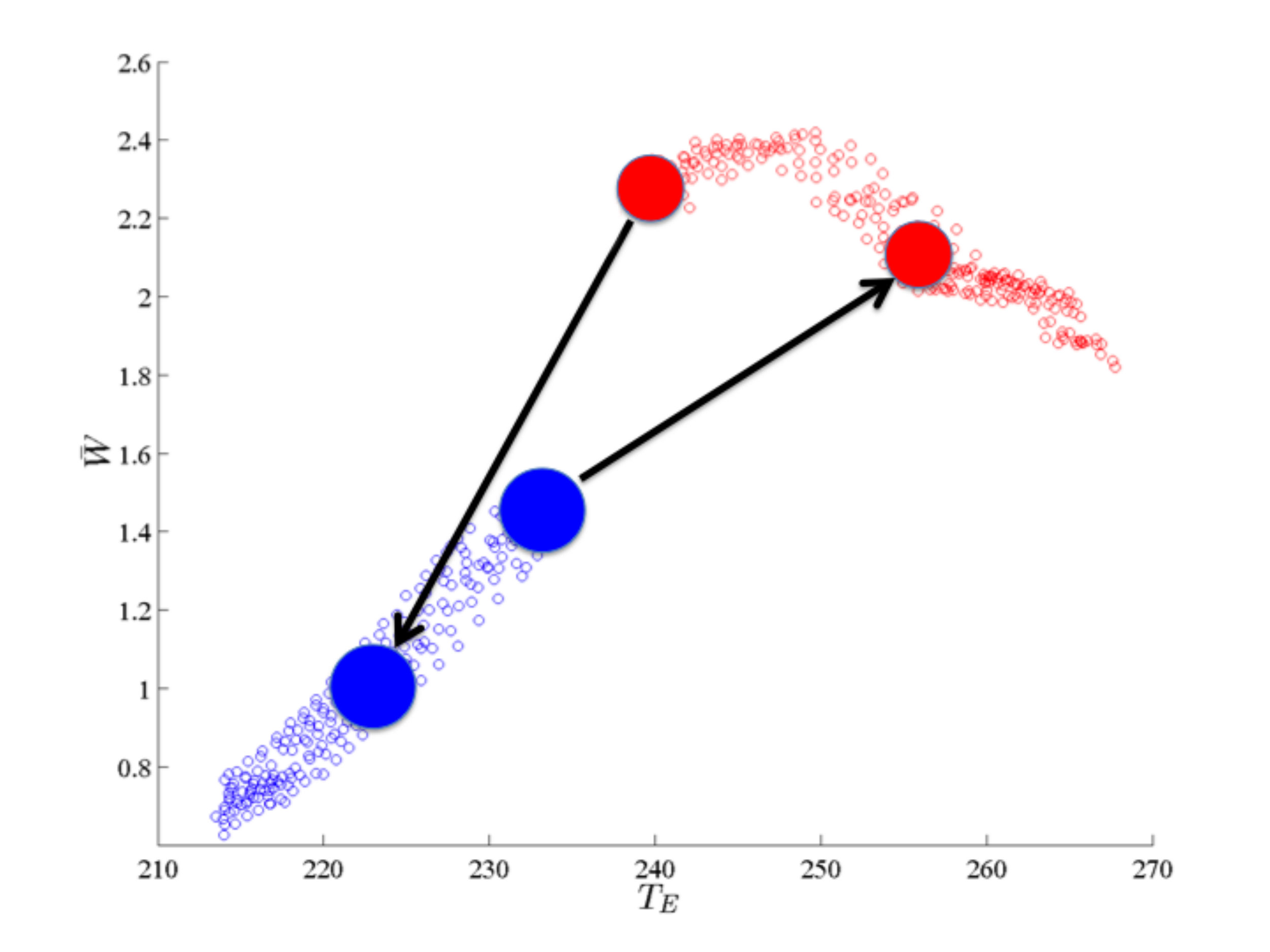}
  \label{dodici_a}
   }
   \subfigure[]{
    \includegraphics[angle=-0, width=0.7\textwidth]{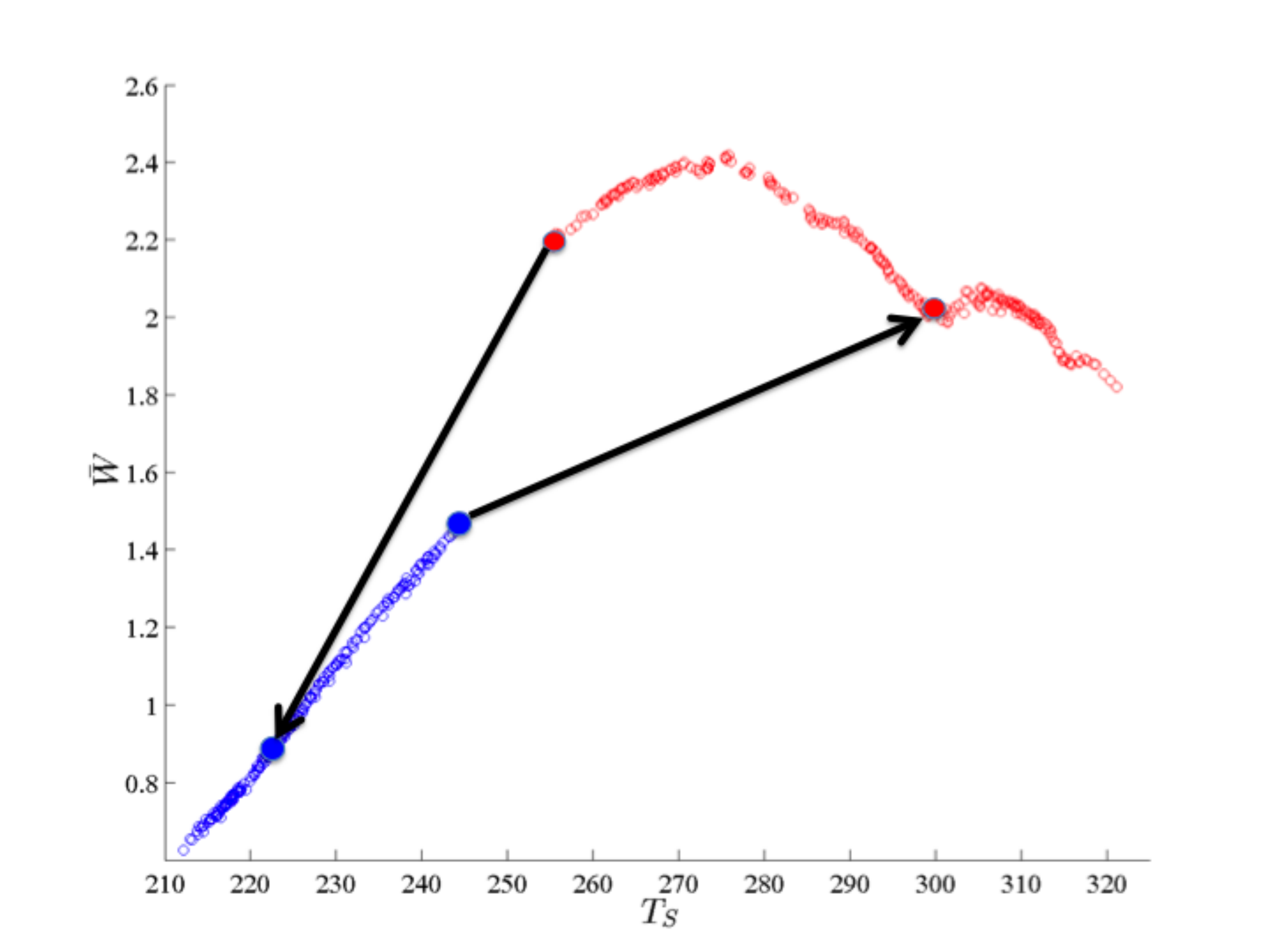}
    \label{dodici_b}
     } 
\caption{  
 \label{dodici}}
\end{figure}

\begin{figure}
\centering
\subfigure[]{
 \includegraphics[angle=-0, width=0.7\textwidth]{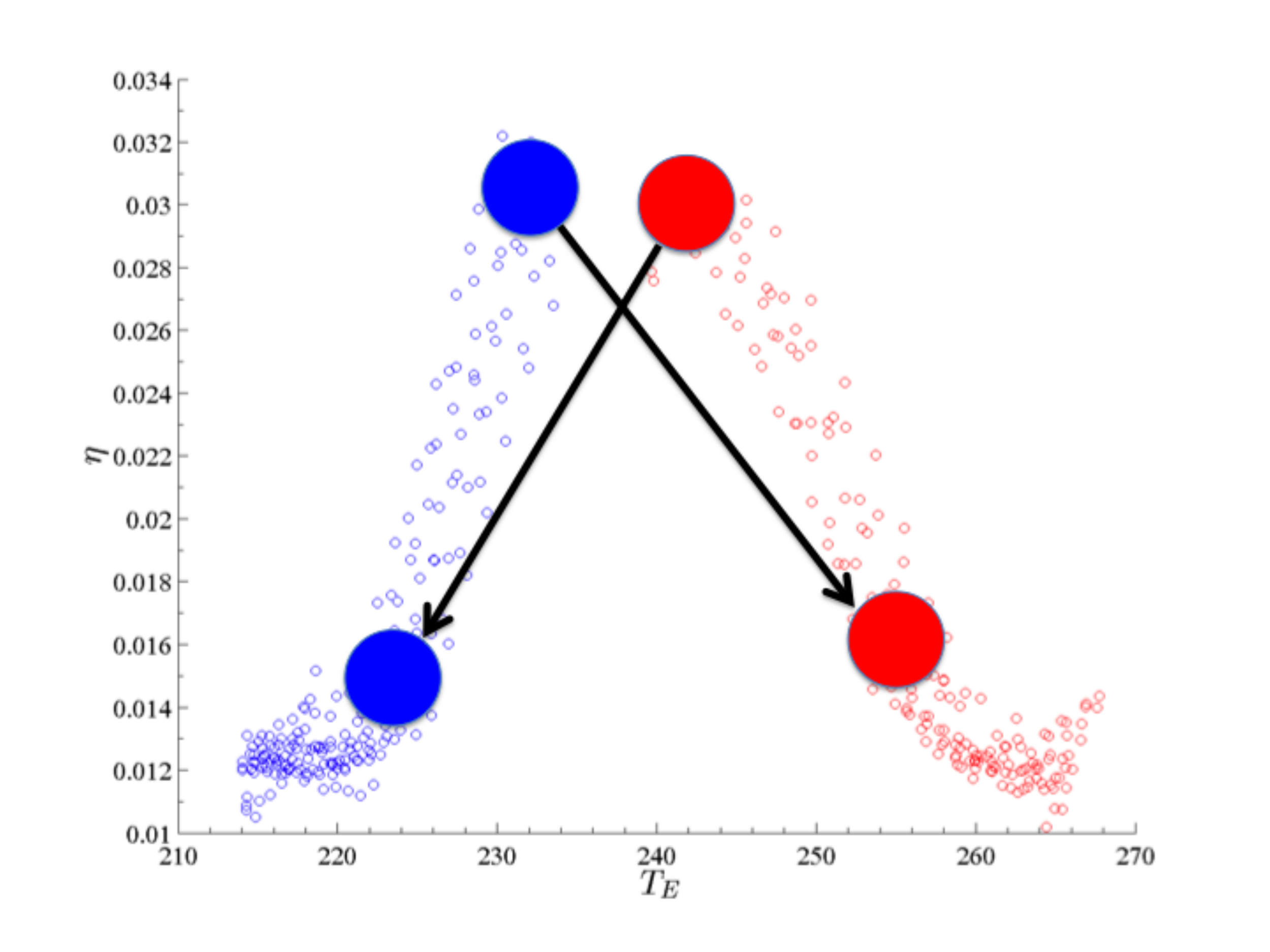}
  \label{tredici_a}
   }
   \subfigure[]{
    \includegraphics[angle=-0, width=0.7\textwidth]{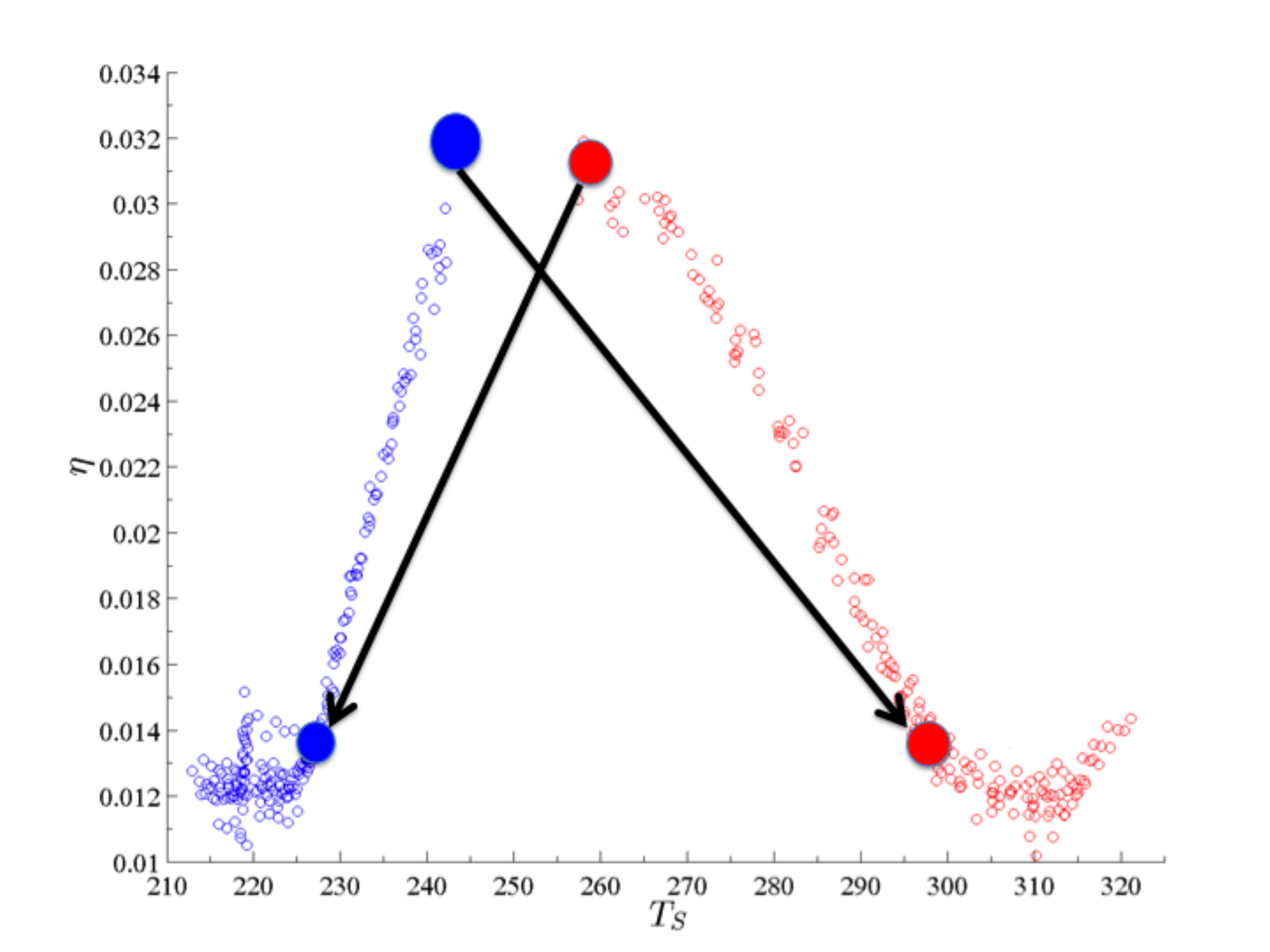}
    \label{tredici_b}
     } 
\caption{ 
 \label{tredici}}
\end{figure}

\begin{figure}
\centering
\subfigure[]{
 \includegraphics[angle=-0, width=0.7\textwidth]{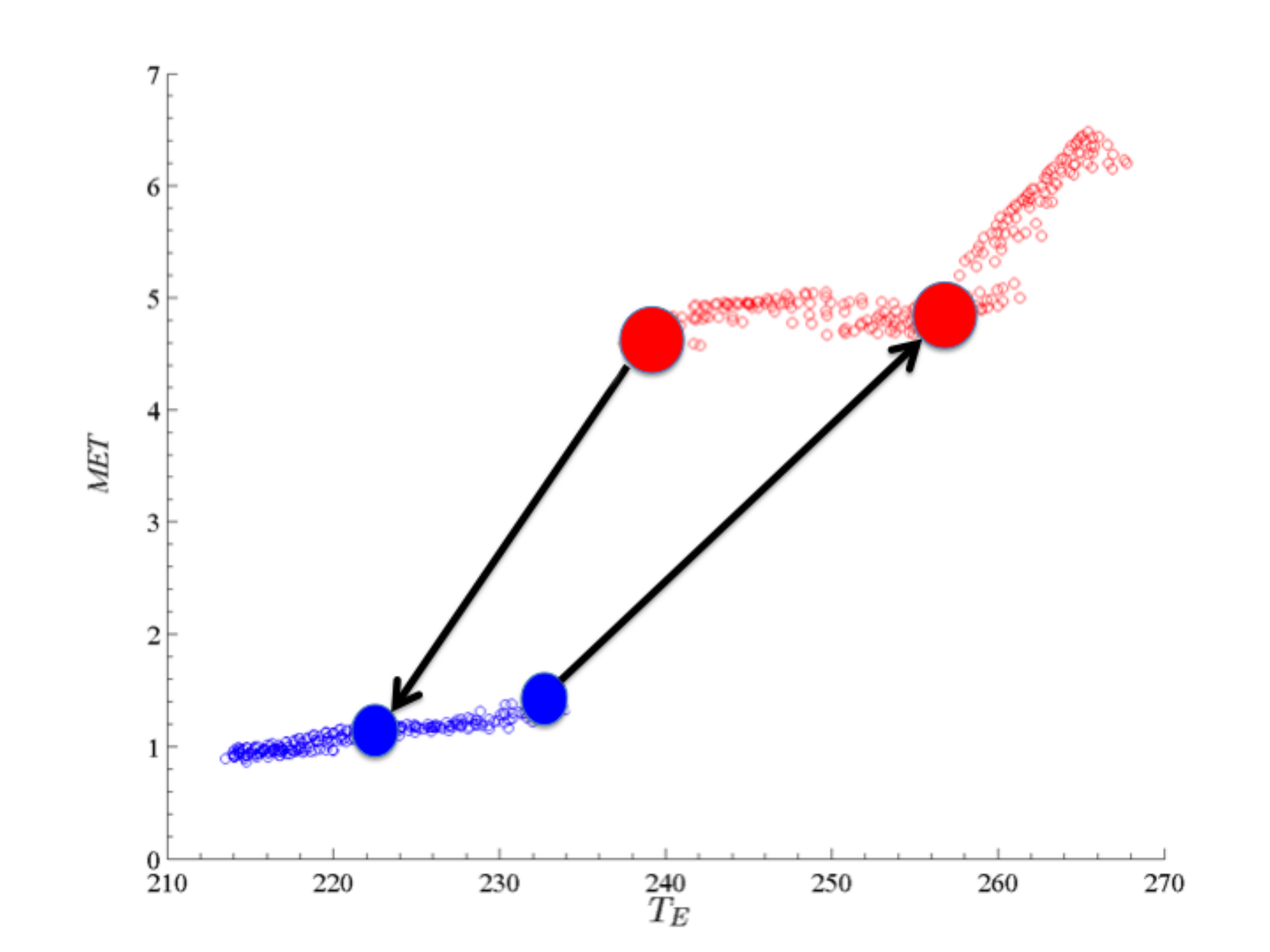}
  \label{quattordici_a}
   }
   \subfigure[]{
    \includegraphics[angle=-0, width=0.7\textwidth]{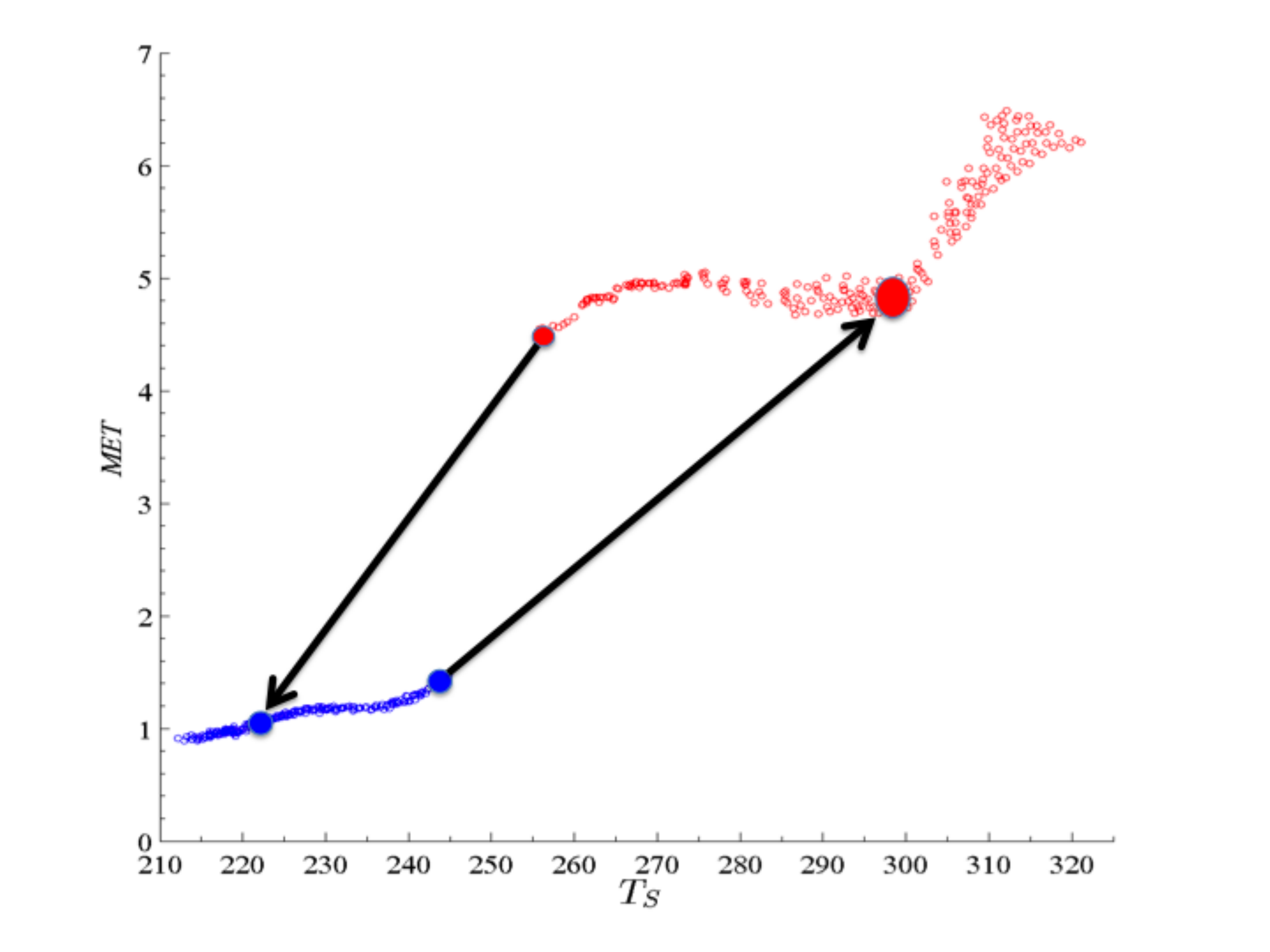}
    \label{quattordici_b}
     } 
\caption{  
 \label{quattordici}}
\end{figure}

\begin{itemize}
\item  Figure \ref{primo} \\(a) Surface temperature and (b) material entropy production for steady states obtained for different values of the solar constant $S^*$. The present climate
 is marked with a black circle, the W states in red and the SB states in blue. Adapted from \citet{LucFr}

\item Figure \ref{secondo}\\    Contour plot of surface temperature (K) as a function of $S^*$ and the [CO$_2$]. The lower SB (a) and upper W  (b) manifolds are shown. 
The transition SB$\rightarrow$W and W$\rightarrow$SB are shown by the upper and lower
 purple lines respectively. The blue dots indicate the values of ($S^*$,[CO$_2$]) for which simulations have been performed.     

\item Figure \ref{terzo}\\ Contour plot of $\overline{\dot{S}_{mat}}$ ($10^{-3}$Wm$^{-2}$K$^{-1}$) as a function of $S^*$ and the [CO$_2$] for the lower SB (a) and upper W (b) manifolds. 

\item Figure \ref{quarto}\\  Contour plot of: (a) meridional energy transport $MET$ (PW, $1$ PW$=10^{15} $ W) and  (b) average rate of dissipation of kinetic energy $\overline{W}$ (W m$^{-2}$) as a function of $S^*$ and the [CO$_2$ ] for the W states.

\item Figure \ref{quinto} \\Contour plot of: (a) meridional temperature gradient (K) and (b)  midlatitude vertical temperature difference (K) as a function of $S^*$ and the [CO$_2$ ] for the W states. 

\item Figure \ref{sesto} \\  Contour plot of the efficiency as a function of $S^*$ and the CO$_2$ concentration for the W states. 

\item Figure \ref{settimo}  \\ Contour plot of: (a) meridional energy  transport $MET$ (PW) and (b) average rate of dissipation of kinetic energy (W m$^{-2}$) as a function of $S^*$ and the [CO$_2$ ] for the SB states

\item Figure \ref{ottavo} \\ Contour plot of (a) the Carnot efficiency $\eta$   and (b)  midlatitude vertical temperature difference (K) as a function of $S^*$ and the [CO$_2$ ] for the SB.

\item Figure \ref{dieci} \\ Surface temperature $T_s$ (K) versus   emission temperature $T_E$ (K) for the W and SB states (each circle represents a simulation). We have marked the  transitions $W\rightarrow SB$ (red dot -- arrow -- blue dot) and $SB\rightarrow W$ (blue dot -- arrow -- red dot). Same convention is used in Figs. \ref{undici}-\ref{quattordici}

\item Figure \ref{undici} \\ Material entropy production $\overline{\dot{S}_{mat}}$ ($10^{-3}$W m$^{-2}$ K$^{-1}$) \emph{vs.} (a) emission temperature $T_E$ (K) and (b) surface temperature $T_s$ (K).

\item Figure \ref{dodici}\\  Lorenz Energy cycle strength $\overline{W}$ (W m$^{-2}$)  \emph{vs.}  (a)   emission temperature $T_E$ (K)  and (b) surface temperature $T_s$ (K).

\item Figure \ref{tredici} \\ Carnot efficienct $\eta$ \emph{vs.}  (a)   emission temperature $T_E$ (K)   and (b) surface temperature $T_s (K)$.

\item Figure \ref{quattordici} \\ Meridional heat transport index $MET$ (in PW) \emph{vs.}  (a)   emission temperature $T_E$  (K)  and (b) surface temperature $T_s$ (K).

\end{itemize}

\clearpage

\clearpage

\end{document}